\documentclass[12pt]{article}

\usepackage{newtxtext,newtxmath}

\usepackage{graphicx}

\usepackage[letterpaper,margin=1in]{geometry}

\usepackage{xcolor}

\renewenvironment{abstract}
	{\quotation}
	{\endquotation}

\date{}

\makeatletter
\renewcommand{\fnum@figure}{\textbf{Figure \thefigure}}
\renewcommand{\fnum@table}{\textbf{Table \thetable}}
\makeatother

\usepackage{scicite}

\usepackage{url}

\def\LAOSTO{LaAlO$_3$/SrTiO$_3$ }

\def\LAO{LaAlO$_3$ }
\newcommand{\up}{\uparrow}
\newcommand{\dn}{\downarrow}

\newcommand{\cop}{c^{\phantom{\dagger}}}
\newcommand{\ket}[1]{\vert #1 \rangle}

\newcommand{\braket}[2]{\langle #1 \vert #2 \rangle}
\newcommand{\braopket}[3]{\langle #1 \vert #2 \vert #3 \rangle}
\newcommand{\expval}[1]{\langle #1 \rangle}
\newcommand{\mx}{m_{x}}
\newcommand{\my}{m_{y}}
\newcommand{\mz}{m_{z}}
\newcommand{\wc}{\omega_{c}}
\newcommand{\wy}{\omega_{y}}
\newcommand{\wz}{\omega_{z}}
\newcommand{\al}{\alpha_{l}}
\newcommand{\av}{\alpha_{v}}

\def\scititle{
	Engineered Chirality of One-Dimensional Nanowires
}
\title{\bfseries \boldmath \scititle}

\author{
    Megan~Briggeman$^{1,2}$,
    Elliott~Mansfield$^{3}$,
    Johannes~Kombe$^{3}$,
    François~Damanet$^{5}$,\and
    Hyungwoo~Lee$^{4}$,
    Yuhe~Tang$^{1,2}$,
    Muqing~Yu$^{1,2}$,
    Sayanwita~Biswas$^{1,2}$,\and
    Jianan~Li$^{1,2}$,
    Mengchen~Huang$^{1,2}$,
    Chang-Beom~Eom$^{4}$,
    Patrick~Irvin$^{1,2}$,\and
    Andrew~J.~Daley$^{6}$,
    Jeremy~Levy$^{1,2\ast}$\and
    \small$^{1}$University of Pittsburgh, Department of Physics and Astronomy, Pittsburgh, PA 15260, USA\and
    \small$^{2}$Pittsburgh Quantum Institute, Pittsburgh, PA 15260, USA\and
    \small$^{3}$Department of Physics, University of Strathclyde, Glasgow G1 1XQ, UK\and
    \small$^{4}$University of Wisconsin-Madison, Department of Materials Science and Engineering, Madison, WI 53706, USA\and
    \small$^{5}$Department of Physics, University of Liège, 4000 Liège, Belgium\and
    \small$^{6}$Department of Physics, University of Oxford, Oxford, United Kingdom\and
    \small$^\ast$Corresponding author. Email: jlevy@pitt.edu
}

\begin{document} 

\maketitle

\begin{abstract} \bfseries \boldmath
The origin and function of chirality in DNA, proteins, and other building blocks of life represent a central question in biology. Observations of spin polarization and magnetization associated with electron transport through chiral molecules, known collectively as the chiral induced spin selectivity (CISS) effect,  suggest that chirality improves electron transfer by inhibiting backscattering. Meanwhile, the role of coherence in the electron transport within chiral nanowires is believed to be important but is challenging to investigate experimentally. Using reconfigurable nanoscale control over conductivity at the \LAOSTO interface, we create chiral electron potentials that explicitly lack mirror symmetry. Quantum transport measurements on these chiral regions that constitute effective nanowires for the electrons reveal oscillatory transmission resonances as a function of both magnetic field and chemical potential. We interpret these resonances as arising from an engineered axial spin-orbit interaction within the chiral region.  The ability to create 1D effective electron waveguides with this specificity and complexity creates new opportunities to test, via analog quantum simulation, theories about the relationship between chirality and spin-polarized electron transport in one-dimensional geometries.
\end{abstract}

\noindent

The relationship between chirality and electron transport has been investigated for two decades, following pioneering work by Namaan and Waldeck  who first demonstrated a chiral induced spin selectivity (CISS) effect in photoemission experiments \cite{Ray1999-eo}.
It has been postulated that molecular chirality produces an axial spin-orbit interaction that locks the electron spin to its momentum, inhibiting back scattering, thus providing a rationale for the prevalence of chirality in essentially all life forms \cite{Michaeli2016-ei}.  
There have been numerous experiments linking molecular chirality to a variety of magnetic effects \cite{Naaman2019-wn, Evers2022-wi}.  
However, there are many open questions regarding the origin of CISS effects, including the importance of the helical radius, pitch, and spin-orbit interactions at the ends of these molecules.  Experiments are usually performed at room temperature where it is difficult to disentangle the role of lattice dynamics.

\begin{figure}[ht]
    \centering
    \includegraphics[width=\linewidth]{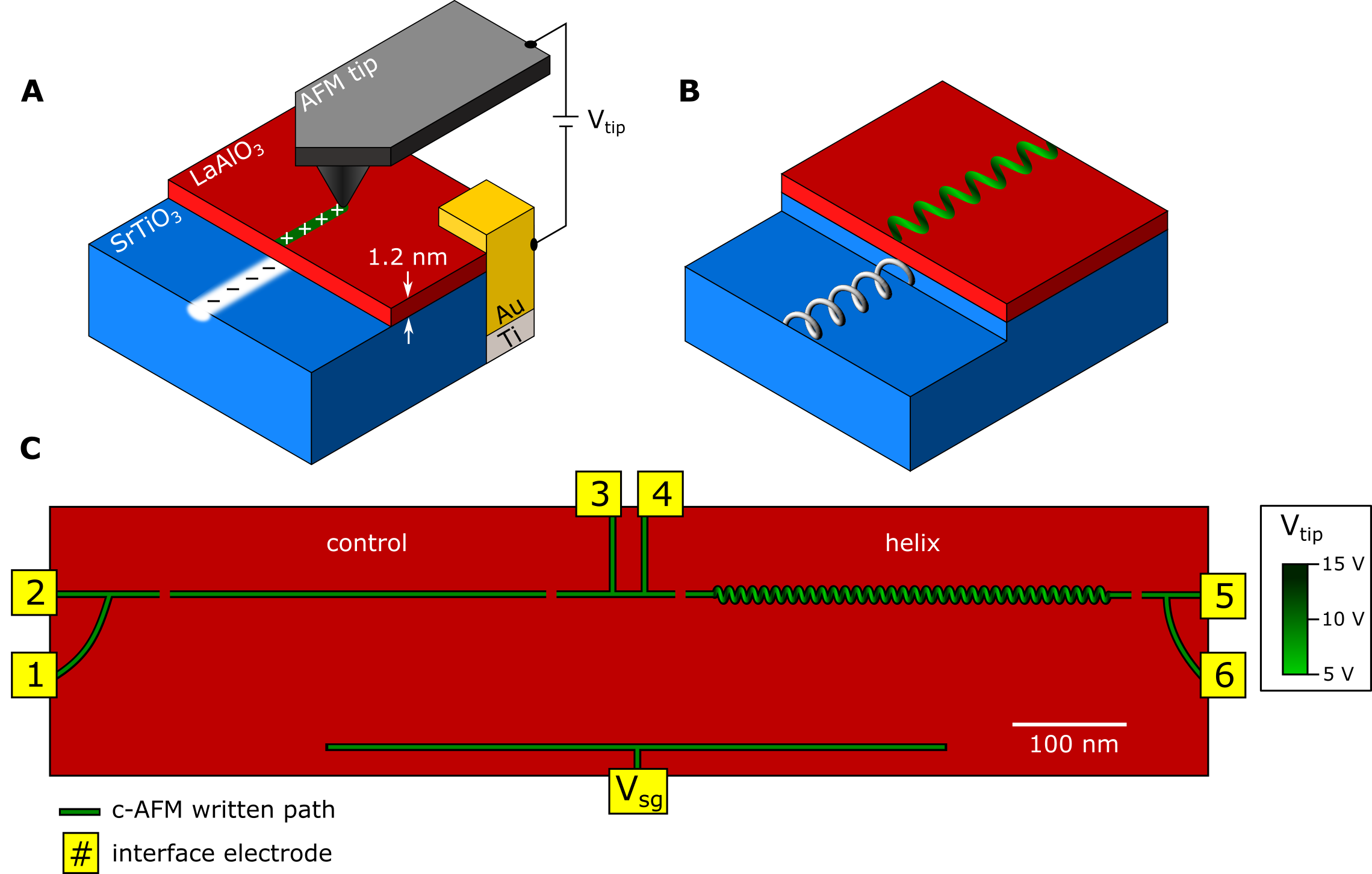}
    
    \caption{Chiral superlattice device.  (A) Conductive atomic force (c-AFM) lithography is used to create conducting channels at the \LAOSTO interface.  (B) Schematic of a chiral conducting channel at the \LAOSTO interface created via vertical and lateral modulations of the tip voltage and tip position respectively.  (C) Schematic of Device A.  The device has two sections: a control waveguide (left) and chiral superlattice (right).  Each section of the device can be probed independently with a 4-terminal measurement, sourcing current through only the section of interest and measuring the voltage drop across that section. White stars indicate highly transparent tunnel barriers.}
    \label{fig:fig1}
\end{figure}

One approach to understanding electron transport in chiral quasi-one-dimensional (quasi-1D) structures involves building and studying them in highly controllable systems, in the sense of analog quantum simulation \cite{Feynman1982-gw, Altman2021-qz}. Quantum simulators based on ultracold atoms in optical potentials have replicated a wide range of phenomena impacting mesoscopic quantum transport \cite{Greiner2002-ox, O-Hara2002-by, Bourdel2004-lb,Singha2011-rl}, including creation of synthetic gauge fields, spin-orbit interactions \cite{Abo-Shaeer2001-vd, Lin2009-iu}, and quantum point contacts with quantized conductance \cite{Krinner2015-dv}.  In our work, we make use of the extreme nanoscale programmability of the metal-insulator transition in \LAOSTO heterostructures, where conductive nanoscale channels can be ``sketched" by scanning a conductive AFM tip over the \LAO surface. A positively biased AFM tip locally switches the \LAOSTO interface from an insulating state to a conductive state (Fig. \ref{fig:fig1}A), while a negatively biased AFM tip locally restores the insulating phase.

One of the remarkable observations for the \LAOSTO system is the high degree of quantization, in contrast to the relatively low mobility $\mu\sim10^{3}$~cm$^2$/Vs found in two dimensions.
The \LAOSTO interface has been shown to exhibit electronic transport that is quantized at or near integer multiples of $e^2/h$, where $e$ is the electron charge and $h$ is the Planck constant, consistent with highly ballistic transport and Landauer quantization.

The \LAOSTO system has already been used to simulate two different types of potentials. One class of devices involves creating a Kronig-Penney-like superlattice, in which the strength of the confinement potential was modulated periodically (period $\lambda$=10 nm) within a channel~\cite{Briggeman2021-ot}, creating a manifold of new subbands and fractional conductance plateaus.
A second class of devices involves serpentine modulation of the electron potential, creating new spin-orbit interactions that shift the subband minima of the engineered effective electron waveguide~\cite{Briggeman2020-wp}.  
Here we combine these two approaches to create quasi-1D potentials that break chiral symmetry within \LAOSTO heterostructures  \cite{Cen2008-ql,Cen2009-of}.  These effective chiral nanowires are created through a combination of two inequivalent perturbations of the ``straight" quantum wire.  The first perturbation involves creating a path for a serpentine potential, given by $y(x)=y_0 +y_k \sin(2\pi x/\lambda)$, where $y_0$, $y_k$ and $\lambda$ are parameters that can be programmed.  The second perturbation involves performing c-AFM lithography with a sinusoidally varying tip voltage: $V_{\mathrm{tip}}(x)=V_0+V_k \sin(2\pi x/\lambda+\phi)$.  The expected impact of the two combined perturbations on the local electron density are lateral and vertical displacements of the wavefunction, with the two displacements shifted by a phase $\phi$.  Choices of $\phi=\pm \pi/2$ rad are expected to yield potentials that break mirror symmetry (Fig. \ref{fig:fig1}B).

A schematic of one of the devices (Device A) is shown in Fig.~\ref{fig:fig1}C.  The device contains a ``helix" potential structure on the right, and a ``control" potential structure on the left.  The helix is created with $y_k$=10 nm, $\lambda$=10 nm, $\phi=\pi/2$.
The control waveguide is straight and written with a constant tip voltage of $V_{\mathrm{tip}}=12~\mathrm{V}$.  Each nanowire is bounded by two nanoscale junctions, created with negative voltage pulses $V_{\mathrm{tip}}=-10~\mathrm{V}$.  
A side gate,  also created using c-AFM lithography, enables an applied voltage $V_{\mathrm{sg}}$ to tune the chemical potential of both the helix and control device. The two electron waveguides can be independently characterized using four-terminal transport measurements, sharing currently only the small section between leads 3 and 4.  The tip position and voltage while writing a chiral superlattice device are shown in Supplemental Fig.~\ref{fig:tip_position}.

\section*{Results}

\begin{figure}[ht]
    \centering
    \includegraphics[width=\linewidth]{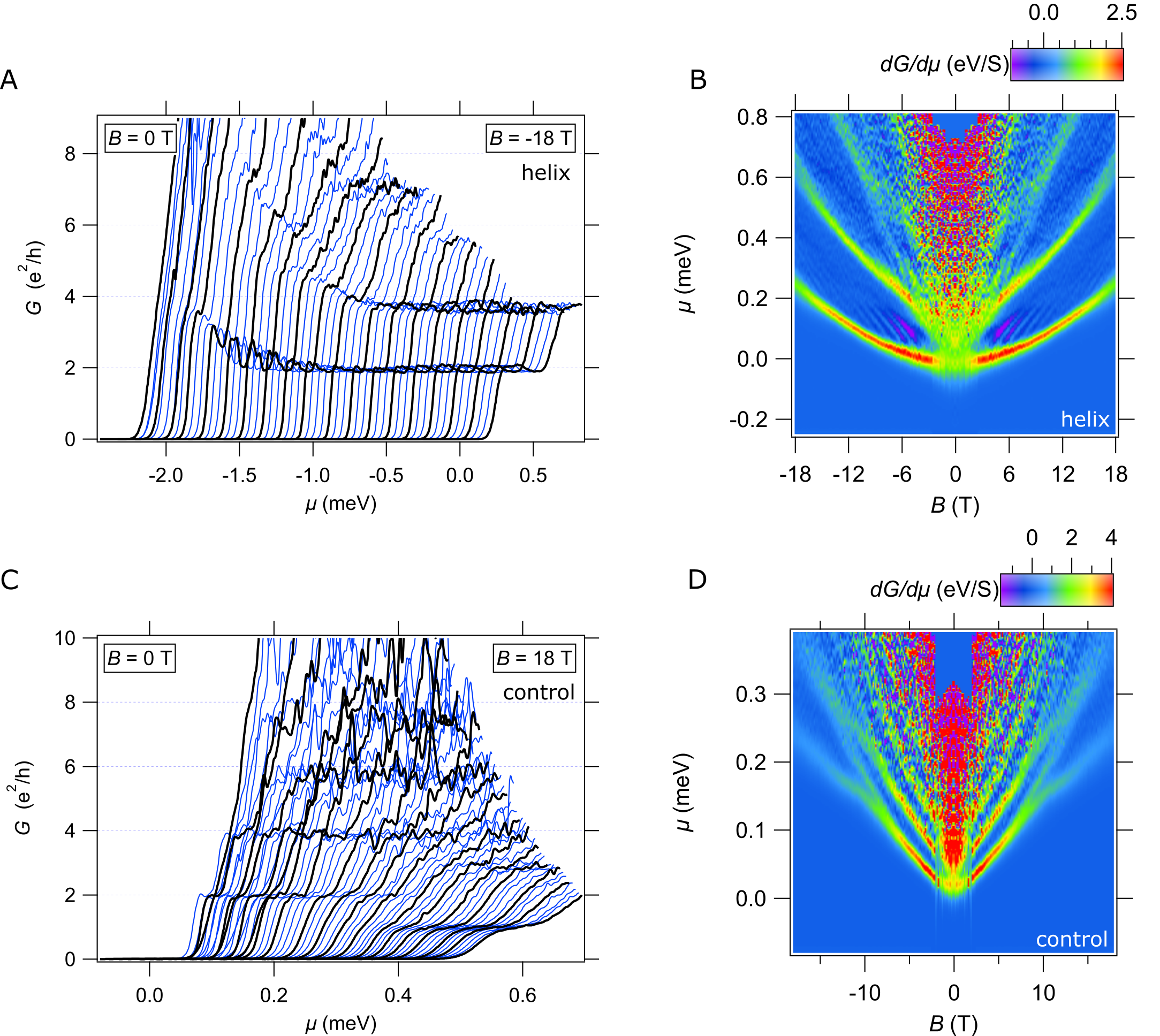}
    
    \caption{Device A transport data at $T=25~\mathrm{mK}$.  (A) Conductance data for the chiral superlattice section of Device A.  Conductance $G$ vs chemical potential $\mu$, each curve is at a different applied magnetic field from $B=0~\mathrm{T}$ to $B=-18~\mathrm{T}$.  Curves at 1 T intervals are highlighted in black.  Curves are offset for clarity. (B) Transconductance $dG/d\mu$ as a function of magnetic field $B$ and chemical potential $\mu$.  Bright (red/yellow) regions indicate increases in the conductance when new subbands become occupied.  Light blue regions are zero transconductance and indicate conductance plateaus.  Dark regions (purple/dark blue) are negative transconductance and indicate decreases in conductance.  (C) Conductance data for the control waveguide section of the device.  (D) Transconductance map of the control section of the waveguide.  There are no oscillations observed in the control section of the device.  It also has a lower pairing field of about $B=8~\mathrm{T}$, where the $2e^2/h$ plateau splits into steps of $1~e^2/h$.  Transconductance data is symmetrized.}
    \label{fig:fig2}
\end{figure}

\noindent Fig. \ref{fig:fig2}A shows the 4-terminal conductance of the chiral Device A as a function of the chemical potential $\mu$, for magnetic fields ranging between 0 T and -18 T. Chemical potential values are calculated by finding the lever arm to convert applied side gate voltage to chemical potential \cite{Annadi2018-az}.  Positive magnetic field values are shown in Fig. \ref{fig:deviceA_positiveB}, positive and negative magnetic field sweeps show similar transport.  Each curve is taken at a different applied magnetic field $B$ from $B=0~\mathrm{T}$ on the left to $B=-18~\mathrm{T}$ on the right, with curves at 1 T intervals hightlighted in black and offset for clarity.  Conductance curves show plateaus at values close to quantized values $G=2e^2/h$ and $G=4e^2/h$ up to $B=-18~\mathrm{T}$.  At low values of magnetic field the $2e^2/h$ plateau is not visible, but there is a feature close to $4e^2/h$.  The transconductance $dG/d\mu$  for the chiral superlattice device (Fig. \ref{fig:fig2}B) is calculated by numerical differentiation of the conductance curves in Fig. \ref{fig:fig2}A.  The bright red/yellow regions correspond to increases in conductance, when new subbands become occupied.  The blue regions correspond to conductance plateaus.  The purple regions correspond to regions of negative differential conductance.  

\begin{figure}[ht]
    \centering
    \includegraphics[width=.8\linewidth]{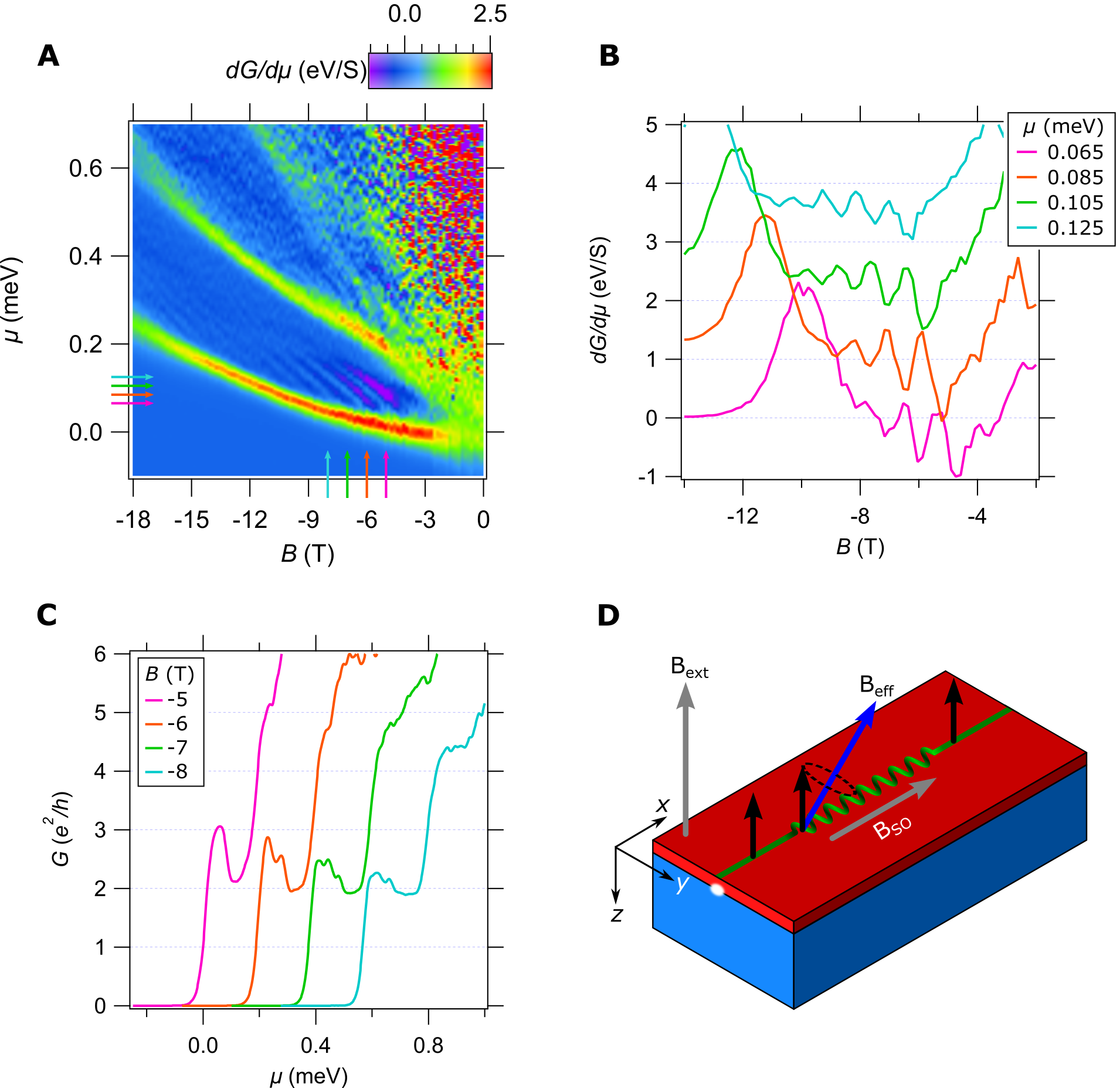}
    
    \caption{Line cuts showing oscillations in Device A.  (A) Transconductance $dG/d\mu$ showing oscillations in the region of the $2e^2/h$ plateau, the region between the lowest subband and the next subband (bright red and yellow regions).  (B) Horizontal line cuts of the transconductance data as a function of magnetic field $B$.  Each curve is taken at increasing values of chemical potential $\mu$, curves are offset for clarity.  (C) Vertical line cuts showing conductance values as a function of $\mu$ at different magnetic field values.  The number of oscillations increase with increasing $B$ field.  The curves overshoot then reduce back to $2e^2/h$.  (D) Schematic of how the chiral superlattice device effects the spin of the electrons in the device.  The applied external field and engineered spin orbit field from the superlattice create a new effective field inside the superlattice.  In the straight portions of the nanowire electrons are polarized in the $z$ direction due to the external $B$ field.  When they enter the superlattice portion of the device they precess around the effective field.  When they exit the superlattice they again will be polarized in the $z$ direction.  If they don’t make a full precession by the time they exit the superlattice the conductance will be suppressed giving rise to the oscillations.}
    \label{fig:fig3}
\end{figure}

The transconductance map reveals characteristic oscillations in the conductance, purple bands in the region above the lowest subband.  The oscillations, oriented at an ``angle" in the $\mu$-$B$ plane, depend on both on the strength of the magnetic field and on the chemical potential.  Line cuts of the transconductance and conductance highlighting these oscillations are shown in Fig. \ref{fig:fig3}B and Fig. \ref{fig:fig3}C.  Vertical conductance line cuts show that the conductance initially rises above $G=2e^2/h$ then falls back down to close to $2e^2/h$.  The number of oscillations increase as the magnetic field is increased.  Fig. \ref{fig:fig3}C at $B=5~\mathrm{T}$ shows one oscillation which increases to 2 oscillations at $B=6~\mathrm{T}$.  The magnitude of the oscillations become suppressed at high magnetic field values, but they are still faintly visible, and even at $B=-18~\mathrm{T}$ the conductance still goes above then back down to $2e^2/h$.  The number of oscillations also increases with increasing values of $\mu$ shown in the horizontal line cuts of the transconductance map in Fig. \ref{fig:fig3}B.  
\vspace{11pt}

The control potential structure of Device A (Fig. \ref{fig:fig2}C,D), an unmodulated straight effective electron waveguide connected in series with the chiral potential, behaves similarly to other published reports \cite{Annadi2018-az,Briggeman2020-cl}. The electron waveguide shows evidence for quantized electron pair transport with $G=2e^2/h$ at zero magnetic field up to a critical magnetic field  $B_P=8~\mathrm{T}$.  Above that critical field, the paired state splits, and the lowest plateau becomes quantized at $G=e^2/h$.  Unlike the chiral device, there are no conductance oscillations found in the control device, and no region of negative transconductance.  
\vspace{11pt}

Finite-bias spectroscopy for the chiral potential (Fig. \ref{fig:1127_IV}A) shows the conductance map as a function of $V_{4\mathrm{T}}$ and side gate $V_{\mathrm{sg}}$ at $B=0~\mathrm{T}$ and the corresponding superconducting peak in the conductance.  Fig. \ref{fig:1127_IV}B shows a typical I-V curve for the chiral superlattice section with a critical current of around 10 nA.  I-V curves for the control device do not show a similar peak, but this could be due to the different back gating conditions for the two sections that were necessary in order to tune the device with the side gate.
 
Fig. \ref{fig:1127_IV}D,F show the finite bias spectroscopy for the conductance and transconductance of the chiral superlattice section respectively.  The transconductance map reveals characteristic diamonds \cite{Glazman1989-yr,Patel1990-qo} which are used to calculate the lever arm and convert gate voltage to chemical potential.

\section*{Discussion}
We have considered a number of single-particle mechanisms to explain the observed electron transport data (Fabry-Perot interference, renormalised g-factor, spin-orbit coupling, multi-orbital or band degeneracies - see Sec. \ref{sec:alternative} of the Supplementary Materials (SM) for details), but they were unable to account for all observed conductance features. However, using effective attractive interactions and spin-orbit coupling between subbands as our basic ingredients, we present below a natural theoretical description that captures (1) enhanced electron pairing and (2) conductance oscillations. 

\paragraph{Enhanced electron pairing.}
The locking of the lowest electronic subbands into a $2e^2/h$ and $4e^2/h$ conductance plateau which persist to higher magnetic fields ($B \sim 18T$) than for unmodulated devices indicates that the underlying pairing interactions between the electrons may be enhanced due to the chiral structure of the potential. 
The microscopic origin of electron-electron interactions in \LAOSTO is still an active subject of research.
As discussed in~\cite{Briggeman2021-ot, Briggeman2020-wp, Damanet2021-lt}, electron pairing could be enhanced due to transverse spin-orbit interactions engineered by the potential modulations. In order to study this effect, we generalized the mean-field theory developed in~\cite{Annadi2018-az, Damanet2021-lt, Briggeman2021-ot, Mikheev2023-iy} (see Sec. \ref{sec:HFB} of the SM for more details) to include both the chiral potential and transverse spin-orbit couplings (SOC), utilizing states characterized by transverse orbital and spin degrees of freedom as a single-particle basis. We find that effective attractive interactions lead to pairing in the lowest two subbands considered here, and that this persists at higher fields when we add the chiral perturbation and the SOC. Moreover, our simulations show the coexistence of singlet and triplet pairs in the chiral region, where the triplet pairs are stabilized by the SOC induced by lateral modulations~\cite{Damanet2021-lt}.

\begin{figure}[ht]
    \centering
    \includegraphics[width=\linewidth]{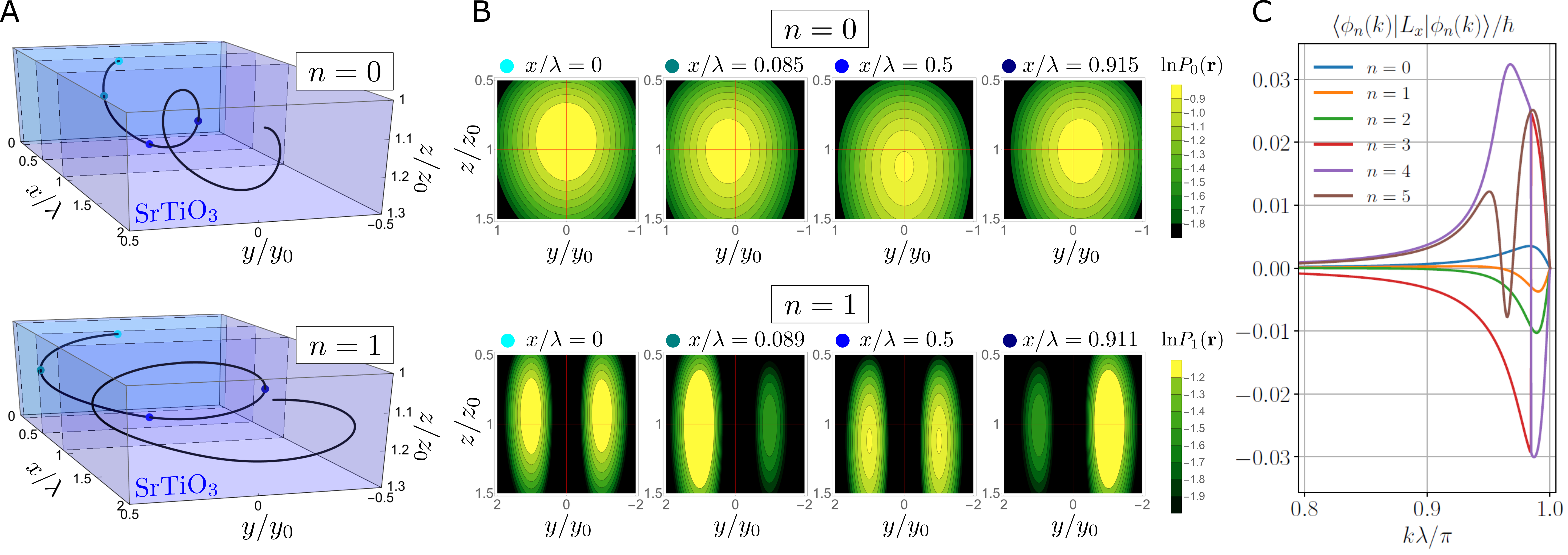}
    \caption{\textbf{(A)}: Trajectory of the center of mass of the electron probability density along the waveguide in the first ($n = 0$, top), and second ($n = 1$, bottom) eigenstate $\ket{\phi_{n}(k)}$ of the trapping potential for a quasimomentum $k = 0.99 \pi/\lambda$. \textbf{(B)}: 2D slices of the previous panel showing the natural logarithm of the normalized probability density $P_{n}(\textbf{r}) = |\braket{\textbf{r}}{\phi_{n}(k)}|^{2}$ in the transverse plane at different positions $x$ for $n = 0$ (top), and $n =1$ (bottom). For both cases, we observe a precession of the probability density. Here $y_{0} = \sqrt{\hslash / (m_{y}\omega_{y})}$ is the characteristic size of the harmonic oscillator in the lateral ($y$) direction (and correspondigly $z_{0}$). \textbf{(C)}: Expectation value of the axial orbital angular momentum for the lowest lying energy eigenstates $\braopket{\phi_{n}(k)}{L_{x}}{\phi_{n}(k)}$ as a function of quasimomentum near the Brillouin zone edge.     Parameters of the numerical simulations are $m_{x} = m_{y} = 1.9m_{e}$, $m_{z} = 6.5m_{e}$, $A=\lambda=10$~nm, $\delta=0.5$, $y_0=26$~nm, and $z_0=8$~nm.}
    \label{fig:theory_HelicalHarmonicOscillator}
\end{figure}

To understand the physical effect of the c-AFM tip modulations on the structure of the single-particle eigenstates in the waveguide, we introduce a chiral harmonic oscillator model which explicitly breaks chiral symmetry (see Sec. \ref{sec:waveguide} of the SM). The effective single-particle physics is captured in Fig. \ref{fig:theory_HelicalHarmonicOscillator} which shows the trajectory of the center of mass (A) and the probability density of the lowest eigenstates (B) in the waveguide as well as their respective axial orbital angular momentum (C). We find eigenstates with non-zero orbital angular momentum $\expval{L_{x}} \neq 0$ (see Sec. \ref{sec:waveguide} of the SM alongside the provided video).
Inside the perturbed region, the electron pairs should thus carry finite $L_{x}$, with their spin degree of freedom generally distributed over the singlet ($S=0$) and triplet ($S=1$) sectors.

\paragraph{Conductance oscillations.} The conductance oscillations shown in Fig. \ref{fig:fig2} take on non-integer values of $e^2/h$ above $2e^2/h$, which suggest that they arise due to a scattering process for pairs of electrons passing through the waveguide.
To account for this behavior, we first note that the modulations of the confining potential give rise to centre-of-mass orbits similar to those of a charged particle in a magnetic field, which are expected to produce an effective axial magnetic field $B_{\mathrm{SO}}$ that arises from an axial spin-orbit coupling \cite{Gutierrez2012-zl}. Fig. \ref{fig:fig3}D shows a schematic of how an engineered axial magnetic field in a finite-length modulated segment, surrounded by unmodulated sections, can produce oscillations in the transport as a function of an applied out-of-plane magnetic field. The effective axial magnetic field, combined with an applied external magnetic field in the $z$ direction, change the spin quantization axis for the charged particles travelling within the chiral waveguide. Particles in the unmodulated portion of the device will be spin polarized in the direction of the applied external magnetic field. As they enter the chiral waveguide, a Rashba-like spin-orbit interaction should cause coherent spin precession around the new quantization axis of the effective $B$ field. The finite length of the device will cause transmission resonances that are periodic in the number of complete precessions. The nature of the oscillations should change depending on the strength of the applied $B$ field, and the energy of the particles, determined by the chemical potential. 

This analysis provides an underlying explanation for the observed conductance oscillations and can be captured in a phenomenological scattering model describing transmission resonances with similar oscillations to those observed in the transconductance data shown in Fig. \ref{fig:fig2}. As the pairs propagate through the waveguide as described above, and shown in Fig. \ref{fig:theory_HelicalHarmonicOscillator}, a spin-orbit interaction of the form $\vec{L} \cdot \vec{S}$ drives coherent oscillations between the singlet and triplet. We model this singlet-triplet interaction by an effective scattering problem for a pseudo spin-$1/2$ particle, and fit the resulting transmission probability to the transconductance fringes (see Sec.~\ref{sec:scattering} of the SM), giving good agreement with the experimental data in Fig.~\ref{fig:fig2}.

\section*{Conclusion}
Programmatic control of chirality in \LAOSTO effective electron waveguides significantly expands the capabilities of this 1D analog quantum simulation platform, enabling the simulation of chiral structures that are believed to be important for spin selectivity in biological systems, donor-acceptor molecules~\cite{Eckvahl2023-xj}, and other manifestations of the CISS effect.  Unlike most naturally occurring systems, the \LAOSTO platform offers a range of mesoscopic building blocks to model not just chirality, but also polarity at the endpoints, which is also believed to contribute to CISS.  While the microscopic origin of  electron pairing in \LAOSTO and aspects of observed ballistic transport are still not well understood, these are well described by phenomenological models.  Further investigation supported by the use of chiral transport will help to lower the ``intellectual entropy" that gives rise to uncertainties in the models used to describe these programmed quantum systems. This will enable chiral 1D systems to serve as building blocks to form 2D superlattices with interesting engineered and/or topogical spin textures.

\clearpage 

%
\bibliographystyle{sciencemag}

%
%


\section*{Acknowledgments}
We acknowledge helpful discussions with Thierry Giamarchi and David Waldeck.
\paragraph*{Funding:}
We acknowledge financial support for this research from AFOSR MURI FA9550-23-1-0368 (J.L.), NSF PHY-1913034 (J.L.), NSF DMR-2225888 (J.L.), Gordon and Betty Moore Foundation’s EPiQS Initiative, Grant 284, GBMF9065 (C.B.E.), and a Vannevar Bush Faculty Fellowship N00014-20-1-2844  (C.B.E). Transport measurement at the University of Wisconsin–Madison was supported by the US Department of Energy (DOE), Office of Science, Office of Basic Energy Sciences (BES), under award number DE-FG02-06ER46327.
\paragraph*{Author contributions:}
Author contributions: HL and C-BE provided samples and room-temperature electrical characterization, MB, YT, MY, SB, Jianan Li, MH, and PI performed experiments.  MB, EM, JK, FD, AD, and JL analyzed data and constructed theoretical models.  All authors contributed to writing the manuscript and/or provided feedback.
\paragraph*{Competing interests:}
There are no competing interests to declare.
\paragraph*{Data and materials availability:}
Data in the figures will be made available on a publicly available repository (e.g., Dataverse).


\subsection*{Supplementary materials}
Materials and Methods\\
Supplementary Text\\
Figs. S1 to S10\\
Movie S1\\
Data S1


\newpage


\renewcommand{\thefigure}{S\arabic{figure}}
\renewcommand{\thetable}{S\arabic{table}}
\renewcommand{\theequation}{S\arabic{equation}}
\renewcommand{\thepage}{S\arabic{page}}
\setcounter{figure}{0}
\setcounter{table}{0}
\setcounter{equation}{0}
\setcounter{page}{1} 


\begin{center}
\section*{Supplementary Materials for\\ \scititle}

Megan~Briggeman,
Elliott~Mansfield,
Johannes~Kombe,
Fran\c{c}ois~Damanet,
Hyungwoo~Lee,
Yuhe~Tang,
Muqing~Yu,
Sayanwita~Biswas,
Jianan~Li,
Mengchen~Huang,
Chang-Beom~Eom,
Patrick~Irvin,
Andrew~J.~Daley,
Jeremy~Levy$^\ast$\\ 
\small$^\ast$Corresponding author. Email: jlevy@pitt.edu

\subsubsection*{This PDF file includes:}
Materials and Methods\\
Supplementary Text\\
Figures S1 to S9\\
Captions for Movies S1\\

\subsubsection*{Other Supplementary Materials for this manuscript:}
Movie S1\\
\end{center}
\newpage


\section{Materials and Methods}
\label{sec:materials}

3.4 unit cell (u.c.) \LAOSTO samples were grown using pulsed laser deposition (PLD) described in more detail elsewhere \cite{Cheng2011-ze}.  Electrical contact was made to the interface by ion milling and depositing Ti/Au electrodes.  C-AFM writing was performed by applying a voltage bias between the AFM tip and the interface, with a 1 G$\mathrm{\Omega}$ resistor in series.  C-AFM writing was performed at room temperature in 30-40\% relative humidity using an asylum MFP3D AFM.  Written samples were then transferred into a dilution refrigerator and cooled to a base temperature of 25 mK.  Magnetic fields up to 18 T were applied in an out-of-plane geometry.  Four-terminal measurements were performed using standard lockin techniques at a reference frequency of 13 Hz, with an applied AC voltage of 100 $\mathrm{\mu}$V. IV curves are also measured by applying a finite DC bias to the device.

The chiral superlattice section of the device (right) is written with a sinusoidal modulation of the tip voltage ($V_0=$ 10 V, $V_k=$ 5 V, $\phi=90^\circ$), and a lateral modulation amplitude ($y_k=$ 5 nm, $2 \pi/k=$ 10 nm), with 34 total periods.  The superlattice is surrounded by two straight segments written with $V_{\mathrm{tip}}=12~\mathrm{V}$, in which highly transparent tunnel barriers \cite{Annadi2018-az} are created by applying negative voltage pulses $V_{\mathrm{tip}}=-10~\mathrm{V}$ to the tip while writing.


\section{Secondary Device}
\label{sec:secondary}

A second Device B (Fig. \ref{fig:deviceB}), shows qualitatively similar features. The chiral potential of this device was written with the same serpentine path parameters, but the vertical modulation was first written with a constant tip voltage $V_{\mathrm{tip}}=10~\mathrm{V}$, and then the path was rewritten with the modulated voltage $V_{\mathrm{tip}}=\pm2.5~\mathrm{V}$.  This two-pass writing process is closer to the way in which the electron waveguide devices are created \cite{Briggeman2020-cl}.  This device also shows oscillations in conductance above the $G=2~e^2/h$ plateau (Fig. \ref{fig:deviceB}D).  Additionally, there are some periodic features in the $we^2/h$ plateau which are visible in Fig. \ref{fig:deviceB}C.  In the conductance these features are due to a periodic change in the slope of the conductance jump from an insulating state to the $2~e^2/h$ plateau.

\section{Waveguide models}
\label{sec:waveguide}
Here we dicuss the structure of the eigenstates of the electrons in the waveguide in more detail. We present two distinct waveguide models that capture the chiral perturbation and yield qualitatively similar single-particles eigenstates and orbital momentum features. The first (more accurate) model accounts for a trapping potential along $z$ that takes into account the fact that the electrons cannot penetrate in the LaAlO$_3$ layer (half-harmonic potential with spatially-dependent frequency). The second model consists in a simplified version where the confinement along $z$ is a full-harmonic potential with a modulated center.

\subsection{Chiral model}

Consider a particle propagating along the axial ($x$) direction, and confined transversally by a laterally and vertically modulated harmonic oscillator potential. The Hamiltonian reads

\begin{equation}
    H = \frac{p_{x}^{2}}{2m_{x}} + \frac{p_{y}^{2}}{2m_{y}} + \frac{1}{2}m_{y}\omega_{y}^{2} \left( y - A\sin[2\pi x/\lambda] \right)^{2} + \frac{p_{z}^{2}}{2m_{z}} + \frac{1}{
    2}m_{z}\omega_{z}^{2}(1 + \delta \cos[2\pi x/\lambda])^{2}z^{2} ~,\quad z\geq 0
\end{equation}

\noindent
where $\lambda$ is the wavelength of the modulation, $A$ is shift in the centre of the lateral harmonic oscillator potential, and $\omega_{y,z}$ the trapping frequencies in the transverse ($y$ and $z$) directions. We emphasize that we are treating the confinement along $z$ as a modulated half-harmonic oscilltor potential (which is infinite for $z<0$). These chiral modulations lead to an effective coupling between the transverse degrees of freedom, such that the Bloch eigenfunctions $\ket{\phi_{n}(k)}$ will be superpositions of different occupations in the two transverse directions $\ket{k, m_{y}, n_{z}}$. 

Fig.~\ref{fig:theory_HelicalHarmonicOscillator} (A-B) shows the probability densities of the ground, and first excited state near the Brillouin zone edge, $\ket{\phi_{0,1}(k\lambda = 0.99\pi)}$. For both states, the centre of mass of the probability density $P_{n}(\mathbf{r}) = |\braket{\mathbf{r}}{\phi_{n}(k)}|^2$ traces a helical path along the axial direction.
The helical undulation means that the eigenstates carry finite axial orbital angular momentum close to the Brillouin zone boundary, $\expval{L_{x}}$, as shown in panel (C).

In our model, these eigenstates constitute the relevant conductance channels in the patterned \LAOSTO waveguide (with an additional spin degree of freedom omitted here for simplicity). Interactions (see the discussion in the main text) will lead to pairing, resulting in $2e^{2}/h$ of conductance from each paired transport channel, and are modelled below as attractive interactions in a mean-field theory. 

\subsection{Helical model}

We also investigated the following simplified waveguide Hamiltonian
\begin{equation}
    H = \frac{p_{x}^{2}}{2m_{x}} + \frac{p_{y}^{2}}{2m_{y}} + \frac{1}{2}m_{y}\omega_{y}^{2} \left( y - A\sin[2\pi x/\lambda] \right)^{2} + \frac{p_{z}^{2}}{2m_{z}} + \frac{1}{
    2}m_{z}\omega_{z}^{2 }\left( z - A\cos[2\pi x/\lambda] \right)^{2} ~,
\end{equation}
where the vertical waveguide perturbation was modelled as a direct shift of the vertical harmonic potential. This model yields a true helical motion of the electron probability density and, as for the previous model, a non-zero orbital momentum along $x$ (not shown). We thus conclude that the exact form of the microscopic model for the waveguide is not important: it is the breaking of the chiral symmetry in both models that yields this universal behaviour.

\section{Hartree-Fock-Bogoliubov Theory}
\label{sec:HFB}
To study the effect of interactions in the modulated waveguide we developed a self-consistent Hartree-Fock-Bogoliubov theory \cite{Annadi2018-az, Briggeman2021-ot, Damanet2021-lt, Mikheev2023-iy} in the presence of periodic potentials (which we note in 1D will not describe superconductivity but will properly predict pairing, which is our focus here). We neglect the weak potential along the axial direction in this minimal model, and treat the electrons as free particles along $x$. The single-particle Hamiltonian in the Landau gauge $\mathbf{A} = (-By, 0, 0)^{T}$ reads

\begin{equation}
    H = \sum_{i=\{x,y,z\}} \frac{\Pi_{i}^2}{2 m_{i}} + \frac{1}{2}m_{y}\omega_{y}(y - y_{x})^{2} + V(z) + A_{z}\cos(Qx) - \frac{1}{2}g\mu_B B \sigma_{z} - \mu + \frac{\av\sigma_{y} - \al\sigma_{z}}{\hslash}\Pi_{x} ~.
    \label{eq:Hwaveguide}
\end{equation}

\noindent
The first three terms describe the spin-degenerate kinetic energy of the electron with canonical momentum operators $\vec{\Pi} = \vec{p} - q \vec{A}$, and effective masses $m_{i}$. $V(z) = \frac{1}{2}\mz\wz^{2}z^{2}$ for $z\geq0$ (else $\infty$) is the half-harmonic oscillator confinement (see also discussion above), and $\omega_{y,z}$ the transverse harmonic oscillator frequencies. The vertical modulation of the waveguide is modelled by a Kronig-Penney potential $A_{z}\cos(Qx)$, whilst the lateral modulation shifts the centre of the harmonic oscillator in the lateral (y) direction, $y_{x} = A_{y}\sin(Qx)$. The next terms describe the lifting of the spin degeneracy due to an external magnetic field $B$ via the Zeeman term, the chemical potential $\mu$, and finally a static Rashba spin-orbit coupling (SOC)~\cite{Damanet2021-lt}. $H_{z}$ is a half-harmonic oscillator whose eigenstates are the odd regular harmonic oscillator states, i.e. $E_{z} = \hslash\wz(2n + 3/2)$ and $\psi(x,y,z) = \psi(x,y)\psi_{n}(z)$ since $[H,H_{z}]=0$. The vertical degree of freedom in this model is fully decoupled from the lateral and axial direction (i.e. $[H, H_{z}] = 0$ while $[H_{x},H_{y}] \neq 0$), but for simplicity we will neglect axial-lateral correlations, and assume the eigenfunctions to be separable in the different directions. As we shall see in the following section this assumption neglects structures in the transverse degrees of freedom, but here we are mainly interested in modelling the effect of electron pairing. 

The resulting single-particle Hamiltonian is periodic in $x$. Its eigenfunctions are Bloch waves along the axial directions, and (half) harmonic oscillator eigenstates in the (vertical) lateral directions. 
Expanding the wavefunction in a plane wave axial basis, $\ket{\psi} = \sum_{n_{y}, n_{z}, \sigma} \sum_{q, G} \psi^{(G)}_{n_{y}, n_{z}, q, \sigma} \ket{q + G, n_{y},n_{z},\sigma}$, where we have restricted the quasimomentum $q$ to the first Brillouin zone, and introduced the reciprocal lattice vector $G = \pm Q, \pm 2Q, \dots$ with $Q = 2\pi/\lambda$, the eigenproblem $H\ket{\psi} = \mu \ket{\psi}$ then takes the compact form

\begin{equation}
\begin{aligned}
    \sum_{G'} \mathbf{\Psi}^{(G) \dagger}_{q} \Big[
    \mathbb{Q}_{-2} ~ \delta_{G',G+2Q} + 
    \mathbb{Q}_{-1}\left(q + G + \frac{Q}{2}\right) ~ \delta_{G',G+Q} + 
    \mathbb{E}^{(G)}_{q} ~ \delta_{G',G} \Big. \\
    \Big. \quad +
    \mathbb{Q}_{+1}\left(q + G - \frac{Q}{2}\right) ~ \delta_{G',G-Q} + 
    \mathbb{Q}_{+2} ~ \delta_{G',G-2Q} 
    \Big] \mathbf{\Psi}^{(G')}_{q} = 0 ~,
\end{aligned}
\end{equation}

\noindent
where we have restricted ourselves to the lowest two subbands, $n_{y}=n_{z}=0$, and $\mathbf{\Psi}^{(G)}_{q} = (\psi^{(G)}_{0,0,q,\up} ~  \psi^{(G)}_{0,0,q,\dn})^{T}$, and thus neglected subband mixing as in~\cite{Damanet2021-lt}, where it was shown to be valid for reasonable waveguide parameters. The $2x2$ block matrices read
\begin{align}
    &\mathbb{E}^{(G)}_{q} = \Big(\frac{\hslash\Omega}{2} + \frac{3\hslash\wz}{2} - \mu \Big) 
    - \frac{1}{2}g\mu_{b}B\sigma_{z} 
    + \frac{1}{2}\my\Big(\frac{\wy\wc}{\Omega}\Big)^{2} \Big(y^{2}_{q+G} + \frac{A^{2}_{y}}{2} \Big)
    ~ \\
    &\quad\quad\quad + ~ (\av\sigma_{y} - \al\sigma_{z}) \Big(\frac{\wy}{\Omega}\Big)^{2} (q+G) \nonumber\\
    &\mathbb{Q}_{\pm 1}(k) = \frac{A_{z}}{2} \pm \frac{i}{2}\my \Big(\frac{\wy\wc}{\Omega}\Big)^{2} y_{k}A_{y}
    ~ \pm ~ \frac{i}{2} (\av\sigma_{y} - \al\sigma_{z}) A_{y} \Big(\frac{eB}{\hslash}\Big) \Big(\frac{\wy}{\Omega}\Big)^{2} \\
    &\mathbb{Q}_{\pm 2} = - \frac{1}{8}\my \Big(\frac{\wy\wc}{\Omega}\Big)^{2} A_{y}^{2} ~,
\end{align}
where $y_{k} = \frac{\hslash k}{eB}$, $\Omega = \sqrt{\wy^{2} + \wc^{2}}$ the renormalized trapping frequency along $y$, and $\wc = e B/\sqrt{\mx\my}$ the cyclotron frequency. To write down a many-particle theory we promote the different Fourier amplitudes to fermionic operators in the usual way, and phenomenologically include attractive interactions to generate pairing at the mean-field level \cite{Annadi2018-az, Briggeman2021-ot, Damanet2021-lt, Mikheev2023-iy}. The interaction is parameterised by $U \equiv U(B) = U_{0} \sqrt{1-\frac{\omega_c^2}{\Omega^2}} = U_{0} \frac{\omega_y}{\Omega}$, and has a phenomenological magnetic field dependence \cite{Damanet2021-lt}, where $U_0$ is a bare interaction strength. This makes $|U|$ a decreasing function of the magnetic field, and the mean-fields independent of this effective scaling. The interacting eigenvalue problem is then solved self-consistently. The eigenvalues and eigenvectors define quasiparticle operators, which are used to update the mean fields at each iteration step until convergence is reached (unless otherwise stated we run the self-consistent convergence procedure until the mean-fields are converged to below $1\%$ relative precision.

Fig.~\ref{fig:theory_HFB_phasediagram} shows the pairing correlations of the mean-field model for the chiral waveguide, and control device (straight). 
Panel (a) shows the singlet pairing field $\Delta_{s}$ as a function of $\mu$ and $B$, where attractive interactions lead to singlet pairing in the two subbands up to some critical field $B_\text{P}$. 
The inclusion of the chiral perturbation and SOC in panel (b) reshapes, and increases the singlet pairing phase in the $(B,\mu)$ space to lower chemical potential values, and up to stronger critical magnetic fields.These findings are consistent with recent studies investigating separately the effect of laterally, and vertically induced SOC~\cite{Briggeman2021-ot, Damanet2021-lt}. Note that for control straight waveguides, such a mean-field analysis has been shown to yield qualitative similar results to DMRG calculations~\cite{Briggeman2020-cl}. 

Additionally, we expect the periodic modulation to open band gaps in the single-particle spectrum, which manifest as a small unpaired (zero-conductance) region in $(B,\mu)$ space, subdividing the singlet pairing region into two. Qualitatively similar features, a fracturing of the transconductance lines due to a Kronig-Penney modulation, have been observed in~\cite{Briggeman2021-ot}. The bottom row shows the triplet pairing field $\Delta_{t}$ for the control device (c), and chiral waveguide (d). As shown in~\cite{Damanet2021-lt}, laterally induced SOC stabilises triplet pairing, leading to the coexistence of stable singlet and triplet pairs in the chiral waveguide with $\Delta_{s/t} \gtrsim k_{B}T$. 

\section{Scattering Model}
\label{sec:scattering}
Here we propose a phenomenological scattering model to account for oscillations in the transmission probability, as the electron pairs propagate through the patterned \LAOSTO waveguide. 

Our interacting simulations showed that spin-orbit coupling can be induced by the chiral modulation and stabilises triplet pairs inside the chiral region, whereas only singlet states are present in the leads. A conductance measurement would therefore only measure the transmitted singlets, $P^{t}_{S=0}$. We expect this quantity to oscillate both as a function of external magnetic field $B$ and particle energy $E$ (tuned by the external chemical potential $\mu$), and for simplicity we assume a perfect effective backscattering of the triplet pairs at the waveguide-lead boundary. We formulate the scattering problem in terms of a pseudo spin-$1/2$ particle (representing the singlet and triplet states of the pairs), entering the waveguide from the left lead, and being transmitted into the right lead with a probability $P^{t}_{S}$, with $S= \{0,1\}$ for singlet/triplet respectively. The chiral waveguide is modelled as an axial spin-orbit coupling interaction for the particle which causes the spin and momentum to lock, and the particle to be subject to an effective axial magnetic field (which implicitly depends on its motional state). Since we hope to gain a qualitative understanding, we are not concerned with quantitative differences in the quantum numbers of a single particle or composite bosonic pairs.

To be concrete, consider two leads, $H_{\rm lead} = \frac{p_{x}^{2}}{2m_{x}} + E_{z}\sigma_{z}$, coupled to a spin-orbit region of length $L$ with Hamiltonian $H_{\rm sys} = \frac{p_{x}^{2}}{2m_{x}} + \frac{\alpha}{\hslash}p_{x}\sigma_{x} + E_{z} \sigma_{z}$, where $\alpha$ denotes the strength of the axial spin-orbit coupling term, and $E_{z} = g\mu_B B/2$ the Zeeman energy.
 
We derive four boundary conditions per spin degree of freedom from the continuity of the wavefunction, and the probability flux at both interfaces $x=0$ and $x=L$. Note that the probability flux is related to the particle velocity operator $v_{x} = \frac{i}{\hslash} [H, x] = \frac{p_{x}}{m_{x}} + \frac{\alpha}{\hslash} \sigma_{x}$ inside the central region (and $\alpha=0$ in the leads). The scattering amplitudes of the leads
are then implicitly related through these boundary conditions, and are rewritten into a scattering problem of incoming and outgoing amplitudes, 
$\vec{O} = S \vec{I}$, related by the scattering matrix $S$. For a given input vector of scattering amplitudes $\vec{I}$,
the outgoing amplitudes can now be computed straightforwardly, and reflection and transmission probabilites derived from the probability flux $j_{\sigma}(x) = \frac{i\hslash}{2m_{x}} \Big[ \psi^{*}_{\sigma}(x) \partial_{x} \psi_{\sigma}(x) - \psi_{\sigma}(x) \partial_{x} \psi^{*}_{\sigma}(x) \Big]$. 

The analytical solution to the scattering problem displays oscillations in both the magnetic field, and chemical potential. A first order perturbative expansion of the transmission probability in $E_{\mathrm{so}}/E_{z}$ (leaving the plane wave factors intact) yields the closed form expression

\begin{equation}
    P^{t} = 1 - \Bigg[(\zeta + \eta)\sin^2\left(\frac{k_{+} + k_{-}}{2}L\right) + (\zeta - \eta)\sin^2\left(\frac{k_{+} - k_{-}}{2}L\right) \Bigg] \frac{E_\mathrm{so}}{E_z} + \mathcal{O}\left[\frac{E_\mathrm{so}^2}{E_z^2}\right]~,
    \label{eq:trans_eq}
\end{equation}

\noindent
where the momenta of the eigenstates in the spin-orbit coupled region are given by $(\hslash k_{\pm})^{2} = 2m_{x} \Big[ E + E_{\mathrm{so}} \mp \sqrt{(E + E_{\mathrm{so}})^{2} + (E_{z}^{2} - E^{2})} \Big]$, $E_{\mathrm{so}} = m_{x}\alpha^2/\hslash^2$ is the spin-orbit coupling energy, and $L$ the size of the waveguide. The prefactors are given by $\zeta = [\xi_\downarrow^2-2 + (\xi_\downarrow^2 - 1) \xi_\uparrow^2]/(2\xi_\downarrow\xi_\uparrow)$ and $\eta = -(\xi_\downarrow^2 + \xi_\uparrow^2)/2$, where $\xi_{\sigma} = k_{\sigma}L_{B}$, $\hslash k_{\sigma} = \sqrt{2m_{x}(E \mp E_{z})}$ are the momenta for the spin-polarised states in the leads, and $L_{B} = \sqrt{\hslash^{2} / (m_{x}E_{z}})$ the magnetic length scale in the system. Since the transmission probability is related to the experimental conductance, we fit $dP^{t}/dE$ to the transconductance fringes, and find consistent qualitative agreement with oscillation periods of $\Delta B \sim 1$T across a range of chemical potentials. 
We find that $\alpha=0.45$meVnm, and $g=0.85$ provide the best estimates for the scattering model parameters, consistent with other studies on \LAOSTO samples \cite{Damanet2021-lt,Caviglia2010-bp}.
We note that there is a functional relationship between $\alpha \lesssim 2$ meVnm and $g \lesssim 2$ which will give rise to similar paterns of maxima and minima in the transmission as these parameters are increased or decreased, which would also be consistent with the transconductance data. For larger $\alpha \gtrsim 2$ meVnm and $g \gtrsim 2$, additional oscillations appear along the fringes, which do not appear in the experimental data.

\section{Alternative models we considered (and discounted)}
In this section, we discuss alternative models we considered to explain the experimental data and why we discounted them.

\label{sec:alternative}
\subsection{Farby-Perot interference}

A question one could naturally ask when looking at the oscillations of the conductance or transconductance data is whether the observed behaviour could be simply explained by the scattering of the particles due to a potential barrier without the need to include spin-orbit coupling as presented in the scattering model section above, nor attractive interactions to explain pairing of particles. This would produce Fabry-Perot-like interferences and oscillations in the transmission. However, this interpretation has two main shortcomings. First, the chiral potential is the combination of a vertical and lateral potentials, which have been investigated individually in previous works~\cite{Briggeman2020-wp, Briggeman2021-ot}. If one accepts the assumption that the chiral potential can be viewed as a potential barrier associated with Fabry-Perot-like interferences, it would be reasonable to consider the same for the vertical and lateral potentials. However, none of them were producing conductance or transconductance oscillations. Second, the observed conductance oscillations exhibit an amplitude of more than $e^2/h$. The most reasonable explanation for such conductance are either the presence of spin-orbit coupling or attractive interactions leading pairs of particles. However, again, attractive interactions alone would not explain why conductance oscillations are not observed in the vertical or lateral potential experiments. Hence, we conclude that the most natural way to undertand the experimental data is to consider that the combination of vertical and lateral modulations leads not only to the effects observed in the individual experiments but also to an additional effect, namely a longitudinal spin-orbit coupling which, together with attractive interactions, yield conductance oscillations of more than $e^2/h$.

\subsection{Renormalised g-factor}

We investigated other possible explanations for the observed subband collapse and re-entrant 'pairing' in the data, without invoking an effective pairing interaction in the system. To this end we explored if a renormalisation of the g-factor of the system within a single-particle model could give rise to qualitatively similar features as the ones observed in previous experiments~\cite{Annadi2018-az,Briggeman2020-cl}. We focus here on two particularly striking features in the transconductance data - $(1)$ the coalescence of two subbands of opposite spin (but same orbital quantum numbers) below some critical magnetic field $B_{p}$, above which we observe a typical Zeeman splitting, and $(2)$ the extended/collapsed (avoided) crossings between subbands of different (same) spin and differing orbital states. Our starting point is a single-particle model which has been shown to capture qualitatively the subbands in these oxide heterostructures very well~\cite{Annadi2018-az,Damanet2021-lt}. To model the 'pairing' features within this single-particle model we considered a variety of functional forms of the g-factor, and found the best agreement using an anomalous, magnetic field dependence of the form $g = g_{\mathrm{eff}}\max\{0, 1 - B_{p}/B\}$~\cite{Mikheev2023-iy}. Here $B_{p}$ takes the role of a critical magnetic field beyond which the Zeeman term becomes dominant, and leads to a splitting of subbands while $g(B) = 0$ mimicks the observed subband collapse. 

Fig.~\ref{fig:gfactor_collapse} shows the conductance map (a) as a function of both chemical potential and external magnetic field (the different subbands are indicated in red, and the functional form of $g(B)$ shown as a white dashed lines), and three linecuts (b) for $g_{\mathrm{eff}} = 0.2$ and $B_{p} = 1T$. At low magnetic fields subbands differing only in their spin quantum number indeed are collapsed onto each other and give the impression of attractive pairing. For $B > B_{p}$ we see that the subbands split, as expected when the Zeeman term becomes the dominant contribution in the Hamiltonian. In this way this simple model seems to be able to capture feature $(1)$ reasonably well. For feature $(2)$ we turn our attention to the crossing of subbands $\ket{0,1,\downarrow}$ with $\ket{1,0,\uparrow}$ at $B \approx 5T$. In previous experiments these bands exhibited re-entrant pairing~\cite{Annadi2018-az,Briggeman2020-cl}, while an avoided crossing was observed between subbands such as $\ket{0,1,\downarrow}$ with $\ket{1,0,\downarrow}$. This is completely absent in this case, we do not observe any effect on a subband due to the presence of an energetically nearby subband. This is intuitively clear by virtue of the fact that this is a single-particle model. Furthermore, `pairing' was observed experimentally between different subbands at different magnetic fields, which cannot be accounted for within this model because the $g(B)$ affects all subbands equally. In principle one could refine this model further by introducing an orbital dependence into the g-factor but  we find this difficult to justify physically, and we concluded that interactions are indeed the most natural way to account for the observed subband collapse, and re-entrant pairing.

In a similar way multi-orbital or band degeneracies could in principle give rise to a subband collapse as observed experimentally in the Pascal conductance series~\cite{Briggeman2020-cl}. However, such degenracies require a carefully fine-tuned relation between the vertical and lateral confinements $l_{y}$ and $l_{z}$, and we have no indication that this confinement relation is satisfied. Also, we can exclude spin degeneracy due to the strong external magnetic fields involved in the experiment.

\subsection{Effect of spin-orbit coupling}

We also investigated the effect of spin-orbit coupling (SOC) on the transport as a potential alternative explanation for the observed features in the experiment. We follow the discussion in~\cite{Damanet2021-lt} by introducing a single-particle model and including two forms of SOC; \emph{lateral} SOC $\sim \alpha_{l}\sigma_{z}$ and \emph{vertical} SOC $\sim \alpha_{v}\sigma_{y}$. Analogously to the previous section we compute the conductance $G$ as a function of chemical potential and external magnetic field, shown in Fig.~\ref{fig:lateralSOC} and Fig.~\ref{fig:verticalSOC} for $\alpha_{l} = 2~$meV$\cdot$nm and $\alpha_{v} = 3$~meV$\cdot$nm respectively.

A lateral SOC does not appear to significantly affect the subband structure, and hence the transport properties. In~\cite{Damanet2021-lt} it was shown that this type of SOC can stabilise triplet pairing in a system with attractive interactions, but no change in the transport properties at the single-particle level was found, in agreement with our modelling here.

Fig.~\ref{fig:verticalSOC} in turn shows the affect of vertical SOC on the transport properties. We can identify very clear deviations from the subband structure in the absence of SOC. Notably the vertical SOC appears to open regions in $(\mu,B)$ parameter space where the conductance jumps up by $2e^{2}/h$ (and subsequently down by $1e^{2}/h$) with increasing chemical potential $\mu$. Whilst we do observe oscillations of the conducatance with an amplitude of $\sim 2e^{2}/h$ it is important to highlight that these oscillations both increase and decrease $G$ in the same way. However, in this model the increase and decrease of $G$ are biased and \emph{not} the same. In stark contrast to the experimental findings we do not observe oscillations of $G$ around a stable conductance plateau with an amplitude of $\sim 2e^{2}/h$. We therefore conclude that this model also does not provide a comprehensive explanation for the observed features, and rather that interations are the most natural and simple explanation which can explain the different observed elements in the transport measurements.

\section{Scattering Length Estimation}
\label{sec:mean}

Here we estimate the scattering lengths for the control and helical devices near the $G=2e^2/h$ plateau, following the method described in Ref.~\cite{Annadi2018-az}.  In that method, we assume a scattering length $L_x=2e^2/h(1-\exp[-l_x/L_x])$, where $x=c$ denotes the control device and $x=h$ denotes the helical device.  We find:

\begin{table}[hbt!]
    \caption{Estimated scattering lengths for control and helical device near $G=2e^2/h$ plateau.}
    \centering
    \begin{tabular}{|c|c|} \hline 
         $L_c$& $L_h$\\ \hline 
         14$\pm$4 $\mu$m&$8\pm3$ $\mu$m \\ \hline
    \end{tabular}

    \label{tab:scattering}
\end{table}


\begin{figure}
    \centering
    \includegraphics[width=5 in]{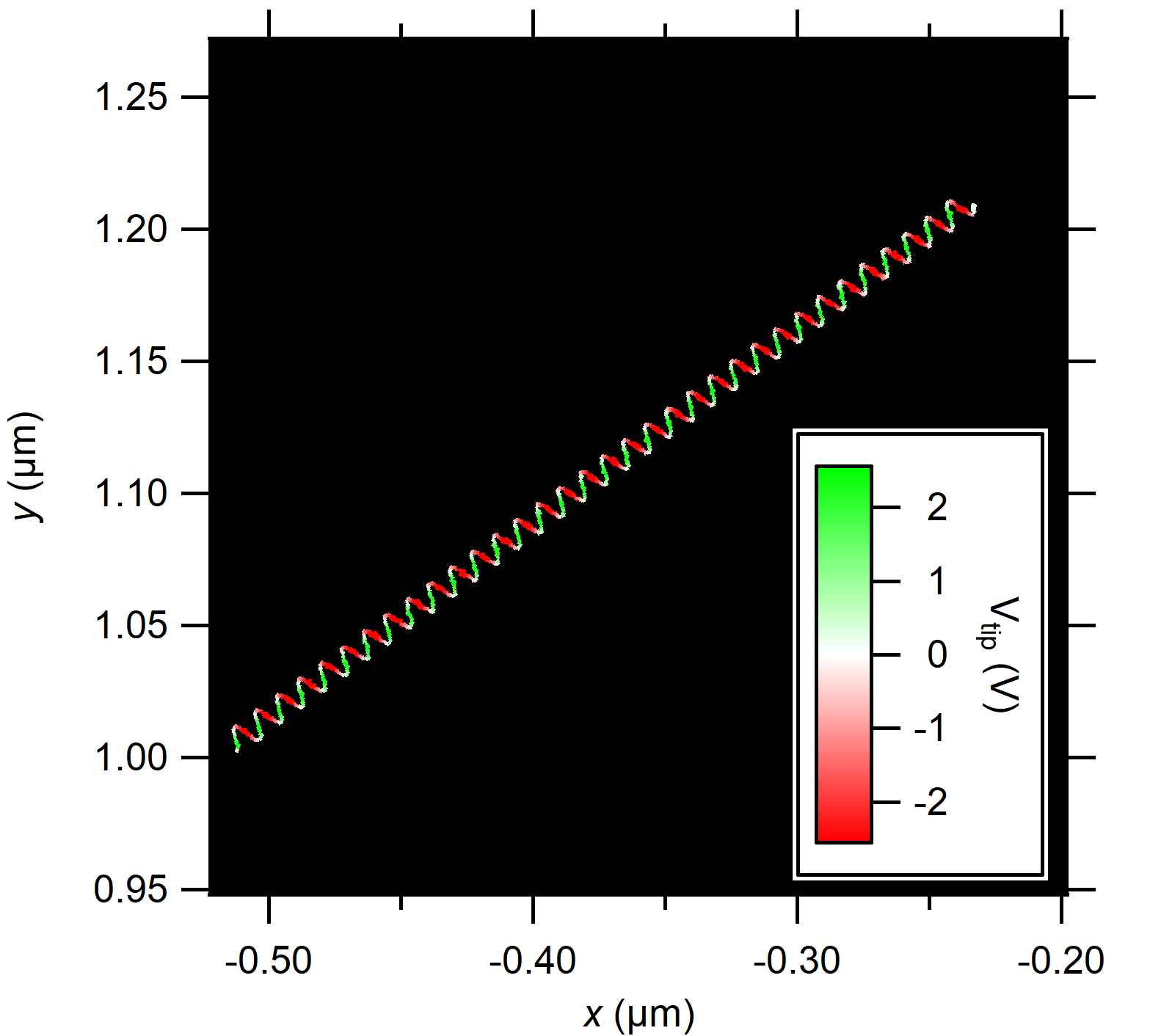}
    
    \caption{AFM tip position and voltage monitored while writing a chiral superlattice section of Device B.  Tip voltage $V_{\mathrm{tip}}$ is represented by the color of the curve.}
    \label{fig:tip_position}
\end{figure}


\begin{figure}[ht]
    \centering
    \includegraphics[width=\linewidth]{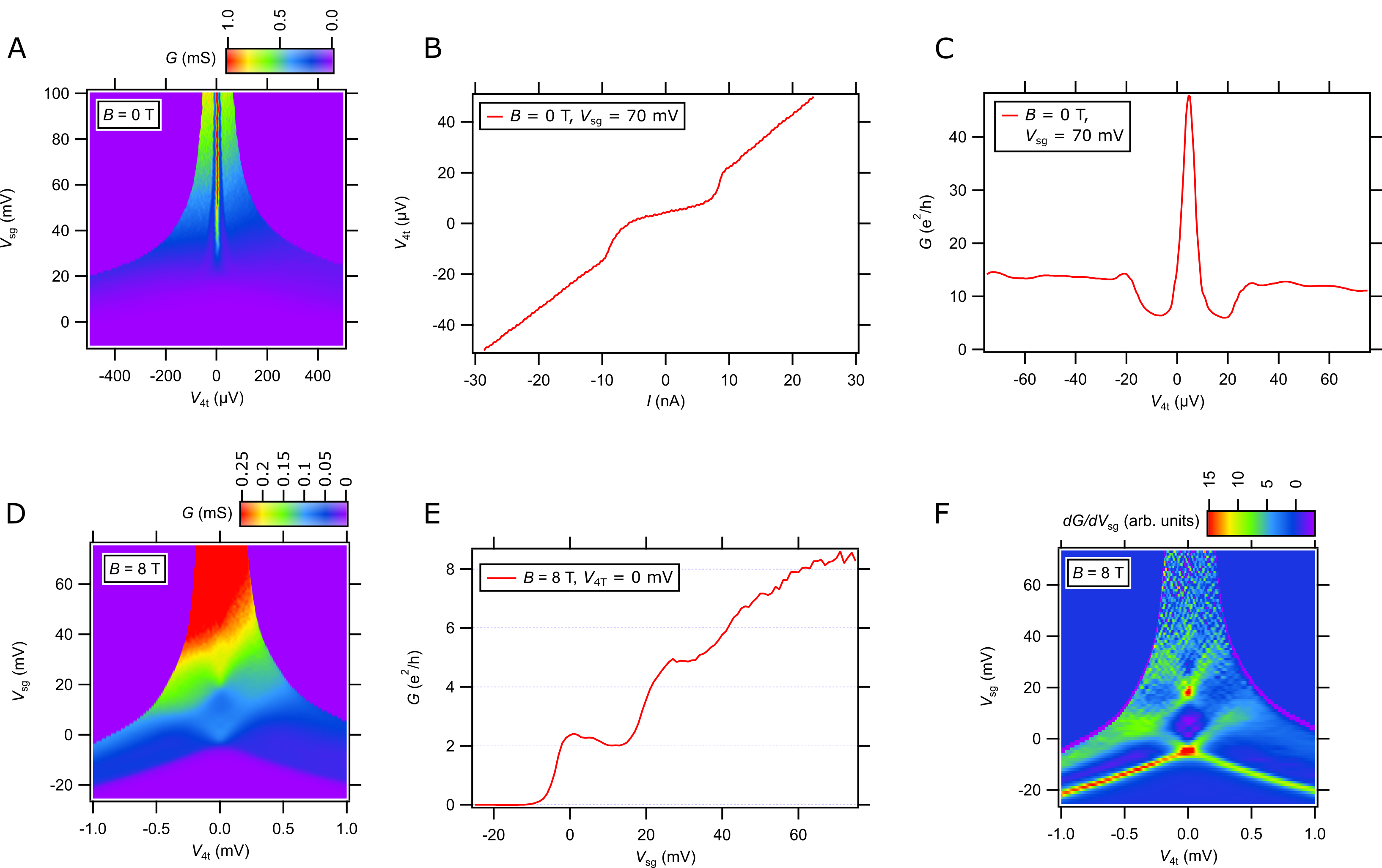}
    
    \caption{Finite bias spectroscopy for the superlattice section of Device A.  (A) IV curve spectroscopy for Device A at $B=0~\mathrm{T}$.  (B) 4-terminal voltage as a function of current for the chiral superlattice section at $B=0~\mathrm{T}$ and $V_{\mathrm{sg}}=70~\mathrm{mV}$ showing the superconducting state with a critical current of around 10 nA.  (C) Horizontal linecut of the conductance map in (A) at $V_{\mathrm{sg}}=70~\mathrm{mV}$ showing the peak in conductance corresponding to the superconducting state.  (D) IV curve spectroscopy at $B=8~\mathrm{T}$.  (E) Vertical line cut at zero bias showing the conductance steps.  (F) Transconductance showing characteristic diamond structure of ballistic transport through the superlattice device.}
    \label{fig:1127_IV}
\end{figure}

\begin{figure}
    \centering
    \includegraphics[width=\linewidth]{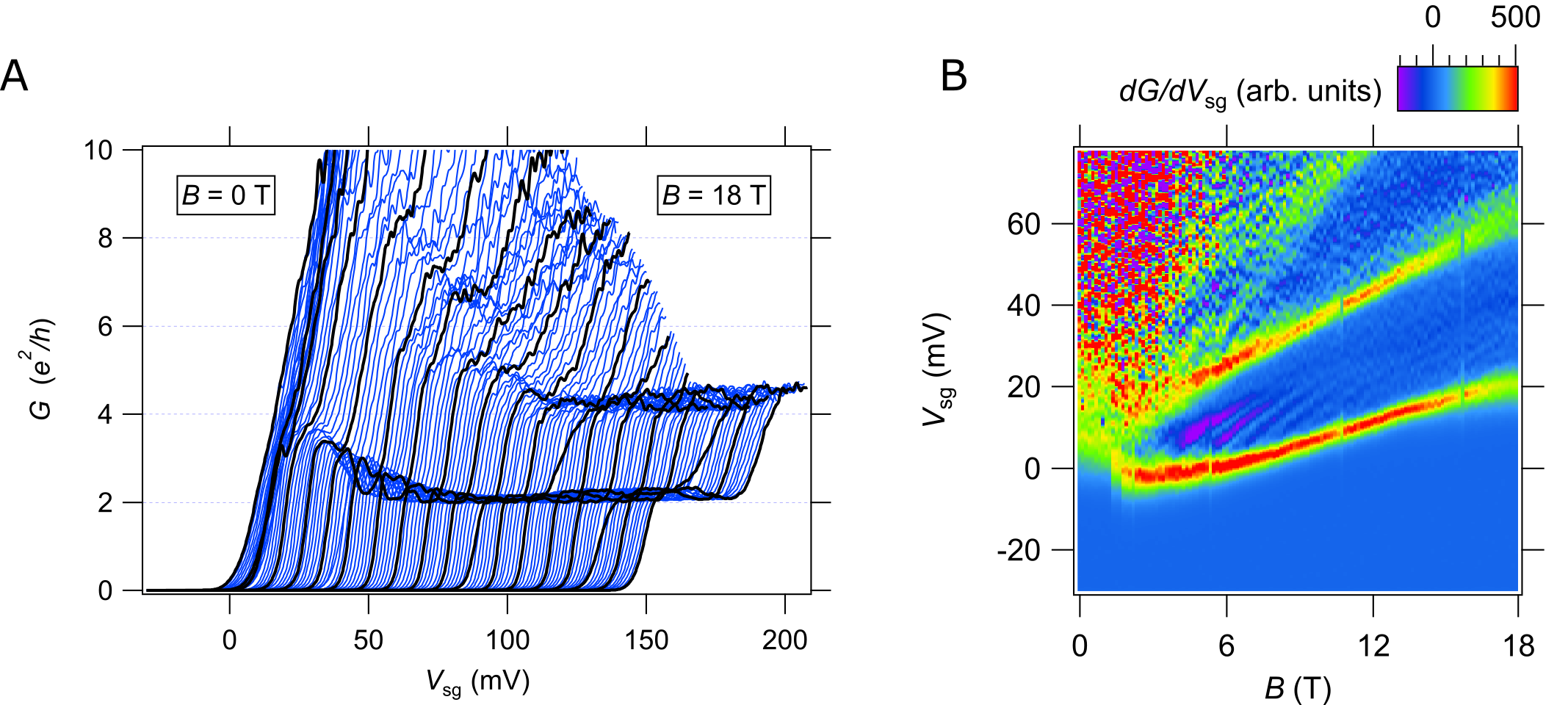}
    
    \caption{Additional data for the superlattice section of Device A.  \textbf{(A)} Conductance curves for the superlattice section of Device A from 0 T to 18 T. \textbf{(B)} Transconductance $dG/dV_{\mathrm{sg}}$ map for positive magnetic field values.  Transport for positive and negative applied magnetic fields is very similar.  During this magnetic field sweep there were several spikes in the temperature of the dilution refrigerator causing distortions in the data at low field, and also periodically at around 5, 11, and 16 T.}
    \label{fig:deviceA_positiveB}
\end{figure}

\begin{figure}[ht]
    \centering
    \includegraphics[width=\linewidth]{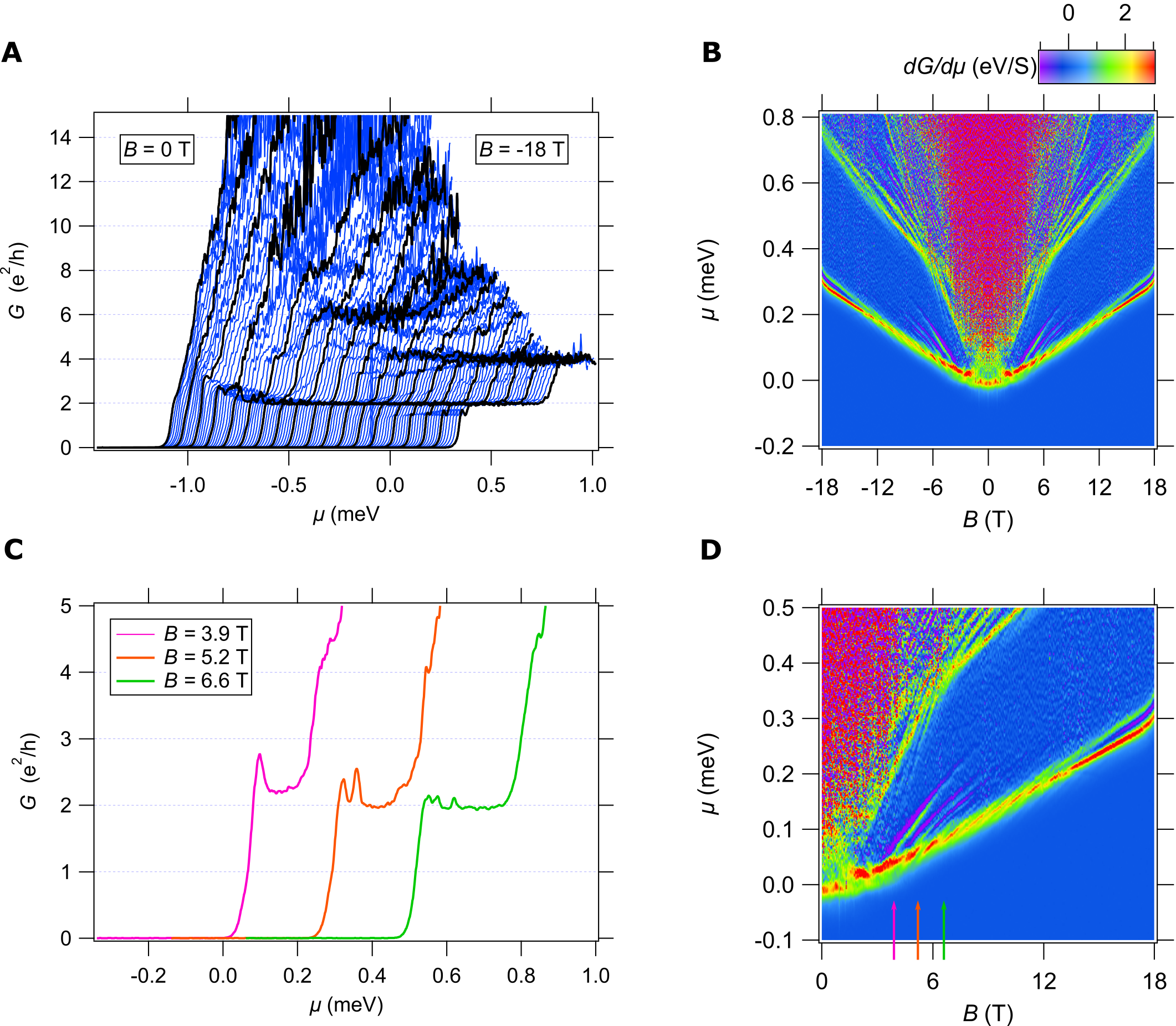}
    
    \caption{Transport data from chiral superlattice Device B. (A) Conductance data as a function of $\mu$, curves are at different magnetic field values from $B=0~\mathrm{T}$ to $B=-18~\mathrm{T}$, curves are offset for clarity.  The value of the first conductance plateau is at $G=2e^2/h$ up to high magnetic field values, although there is a feature that appears at around $B=15~\mathrm{T}$.   (B) Transconductance $dG/d\mu$ as a function of magnetic field $B$ and chemical potential $\mu$.  Data is symmetrized.   (C) Conductance line cuts at three magnetic field values showing the oscillations with a base conductance values of $G=2e^2/h$.  The number of oscillations increases with increasing B field.  (D) Zoom in of the transconductance showing the oscillations and the lowest subband.  Oscillations also appear in the lowest subband, where the conductance is increasing (bright regions) as well as on the plateau (dark regions).}
    \label{fig:deviceB}
\end{figure}

\begin{figure}[ht]
    \centering
    \includegraphics[width=0.5\linewidth]{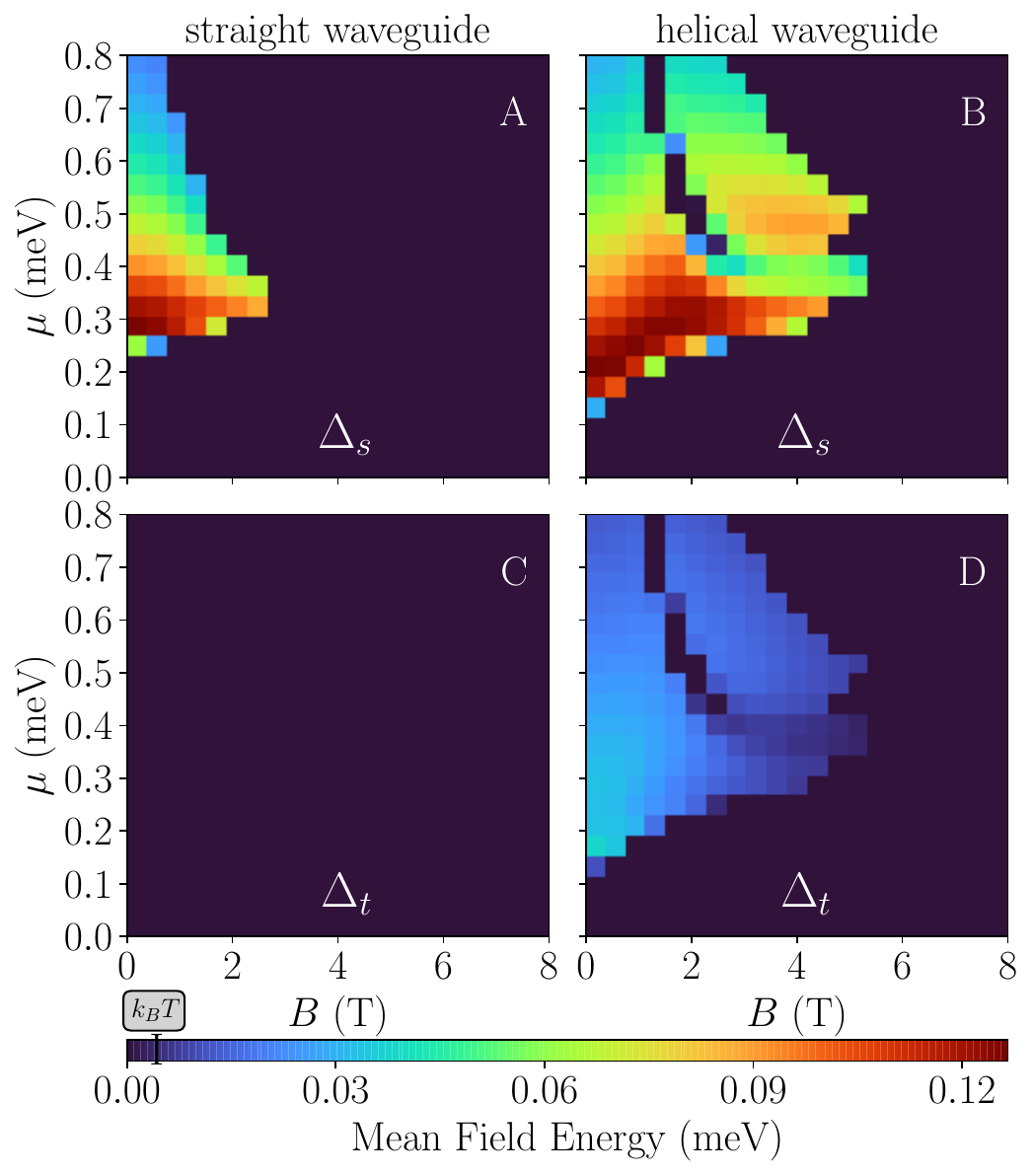}
    \caption{Singlet pairing fields $\Delta_{s} = \frac{U}{2\pi} \int \mathrm{d}k ~ s_{k}$, with $s_{k} = \expval{\cop_{-k,\dn}\cop_{k,\up} - \cop_{-k,\up}\cop_{k,\dn}}/2$ are shown for the \textbf{(A)} straight, and \textbf{(B)} chiral waveguide in the top row, while the bottom row depicts the triplet pairing field $\Delta_{t} = \frac{U}{2\pi} \int \mathrm{d}k ~ |t_{k}|$, with $t_{k} = \expval{\cop_{-k,\dn}\cop_{k,\up} + \cop_{-k,\up}\cop_{k,\dn}}/2$ in the \textbf{(C)} straight, and \textbf{(D)} chiral waveguide. Generally, a pairing field of $|\Delta| \gtrsim k_{B}T \approx 2.15$ $\mu$eV indicates a paired phase associated with $2\frac{e^{2}}{h}$ conductance quanta.
    Here $\cop_{k,\sigma}$ is the fermionic annihilation operator for a subband electron of spin $\sigma$, and $U$ is the attractive interaction strength.
    All simulations were performed for $m_{x} = m_{y} = 1.9m_{e}$, $m_{z} = 6.5m_{e}$, $g = 0.6$, $U_{0} = -5.0$~meV$\cdot$nm, $y_{0} = 26$ nm, $z_{0} = 8.1$~nm, and $T=25$~mK. For the chiral waveguide we additionally used $A_{y}=\lambda=10$~nm, $A_{z}=0.2$~meV, and $\alpha_{v}=\alpha_{l}=2.0$~meV$\cdot$nm. The curved region of reduced pairing in the chiral waveguide is a direct result of the band gap introduced by the periodic modulation. Throughout, we have ensured the mean-fields are converged to a relative precision of $1\%$.}
    \label{fig:theory_HFB_phasediagram}
\end{figure}

\begin{figure}
    \centering
    \includegraphics[width=0.65\linewidth]{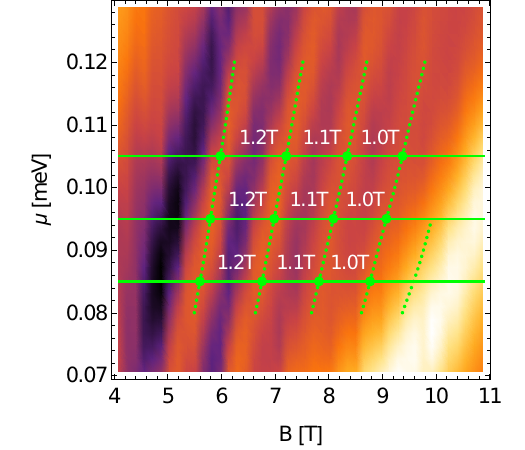}
    \caption{Fits to the experimental transconductance data. We have fitted the scattering model to the location of the fringes  shown in Fig.~\ref{fig:fig2}. Maxima of the fitted scattering model are shown as green dots on top of the experimental data (contour background). The fitting parameters used are $\alpha=0.45$ meVnm and $g=0.85$. We find good qualitative agreement with the experimental data, and observe consistent magnetic field oscillation periods of $\Delta B\sim 1$T. These are shown in white between each fringe for three different chemical potentials.}
    \label{fig:fitting-exp-data}
\end{figure}

\begin{figure}[ht]
    \centering
    \includegraphics[width=1.0\linewidth, trim={30 30 60 60}, clip]{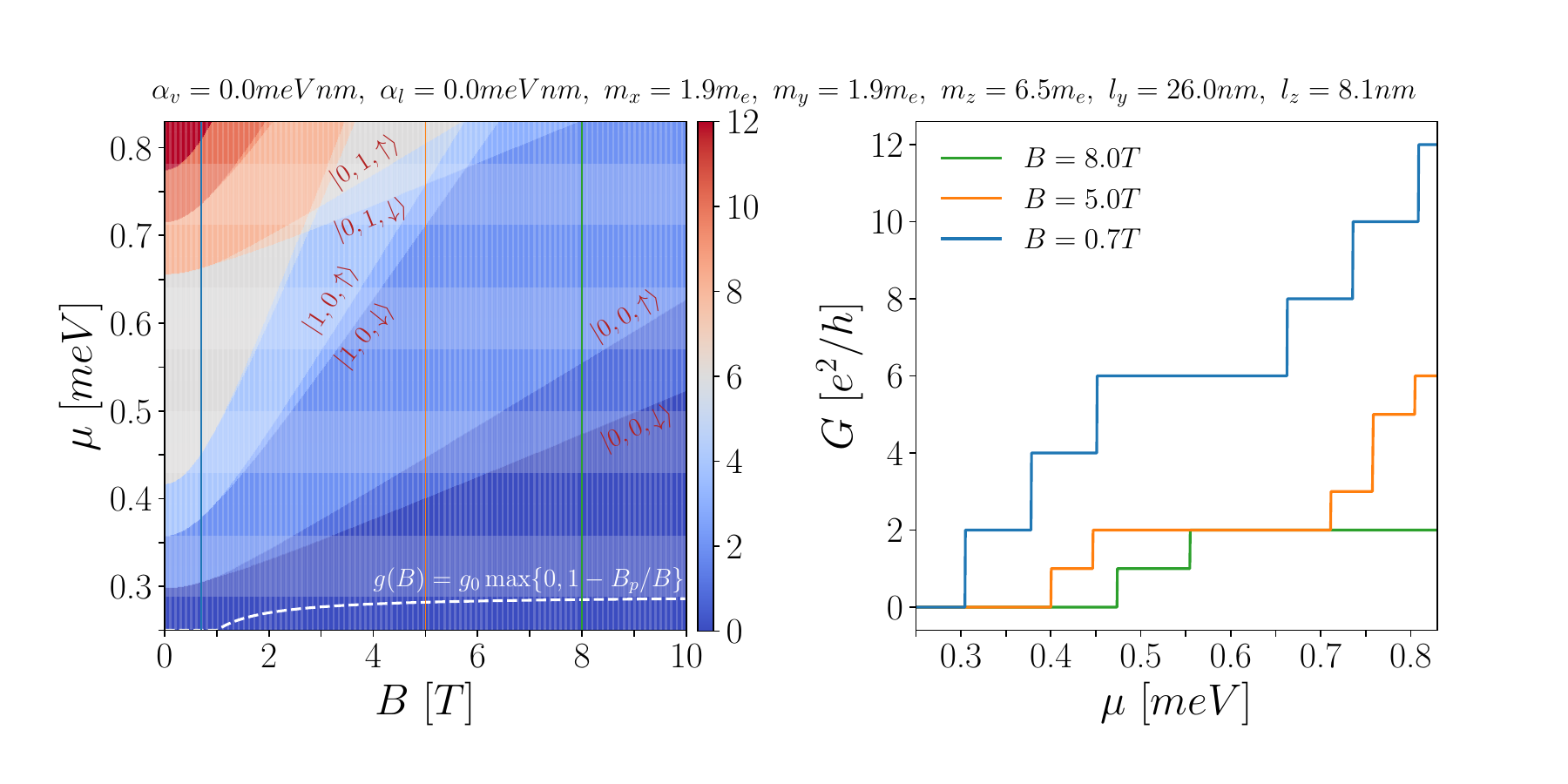}
    
    \caption{(left) Zero-bias conductance $G$ (in units of $e^2/h$) as a function of magnetic field $B$ and chemical potential $\mu$ for $\alpha_v = \alpha_l = 0.0$ meV$\,$nm. The dashed white line indicates the renormalized $g$-factor as a guide to the eye, while the three solid vertical lines show the values of $B$ chosen to plot the conductance as a function of $\mu$ in three distinct regimes shown in the right panel. (right) Conductance as a function of $\mu$ for $B = 0.7$\,T (blue), $B = 5$\,T (orange) and $B = 8$\,T. Other parameters are $m_{x} = m_{y} = 1.9m_{e}$ and $m_{z} = 6.5 m_{e}$ where $m_e$ is the electron mass, $l_y = \sqrt{\hslash/(m_y \omega_y)} = 26$ nm, and $l_z = \sqrt{\hslash/(m_z \omega_z)} = 8.1$ nm.}
    \label{fig:gfactor_collapse}
\end{figure}

\begin{figure}[ht]
    \centering
    \includegraphics[width=1.0\linewidth, trim={30 30 60 60}, clip]{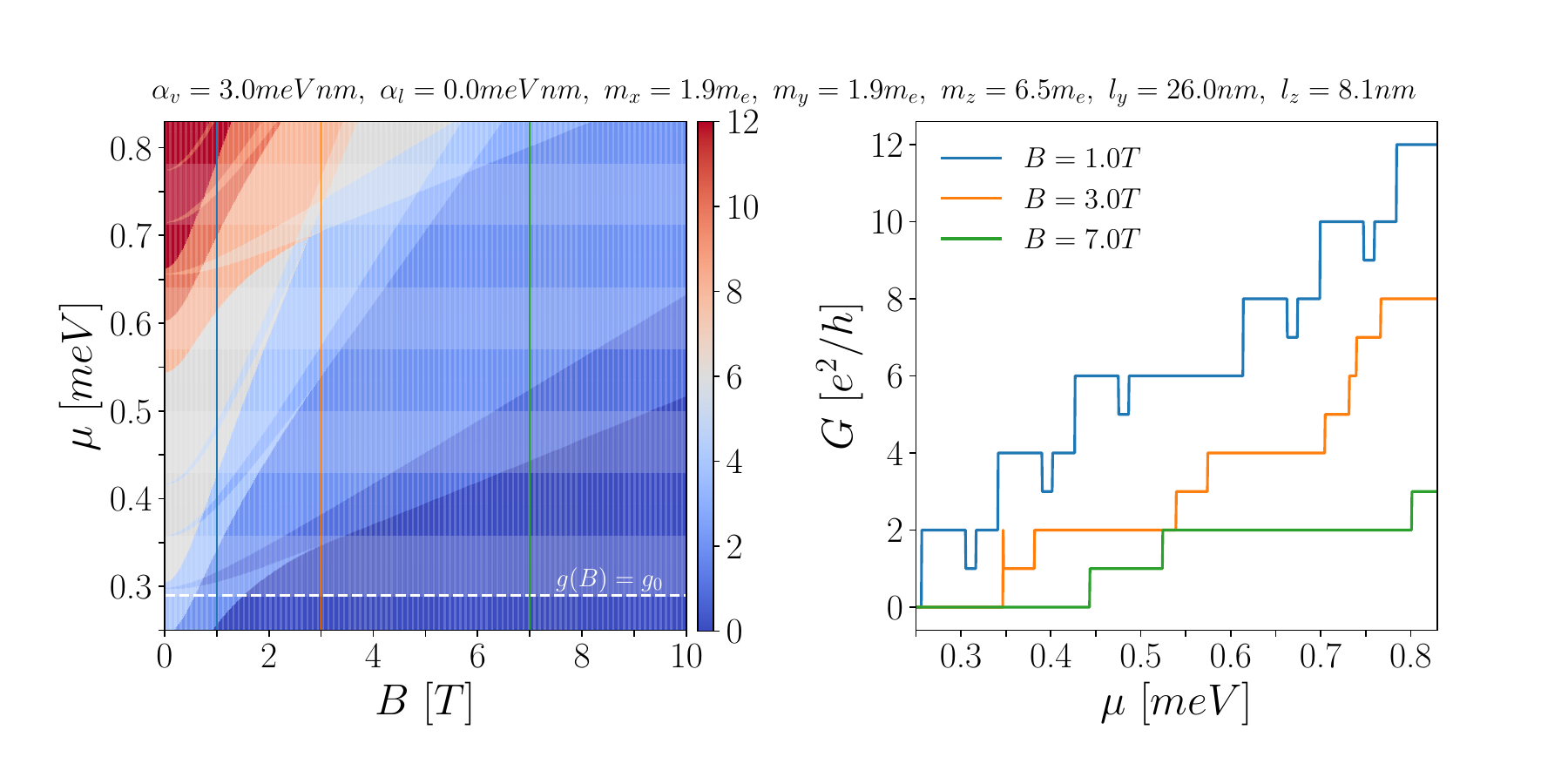}
    
    \caption{(left) Zero-bias conductance $G$ (in units of $e^2/h$) as a function of magnetic field $B$ and chemical potential $\mu$ for $\alpha_v = 3.0$ meV$\,$nm, and $\alpha_l = 0.0$ meV$\,$nm. The dashed white line indicates the renormalized $g$-factor as a guide to the eye, while the three solid vertical lines show the values of $B$ chosen to plot the conductance as a function of $\mu$ in three distinct regimes shown in the right panel. (right) Conductance as a function of $\mu$ for $B = 0.7$\,T (blue), $B = 5$\,T (orange) and $B = 8$\,T. Other parameters are $m_{x} = m_{y} = 1.9m_{e}$ and $m_{z} = 6.5 m_{e}$ where $m_e$ is the electron mass, $l_y = \sqrt{\hslash/(m_y \omega_y)} = 26$ nm, and $l_z = \sqrt{\hslash/(m_z \omega_z)} = 8.1$ nm.}
    \label{fig:verticalSOC}
\end{figure}

\begin{figure}[ht]
    \centering
    \includegraphics[width=1.0\linewidth, trim={30 30 60 60}, clip]{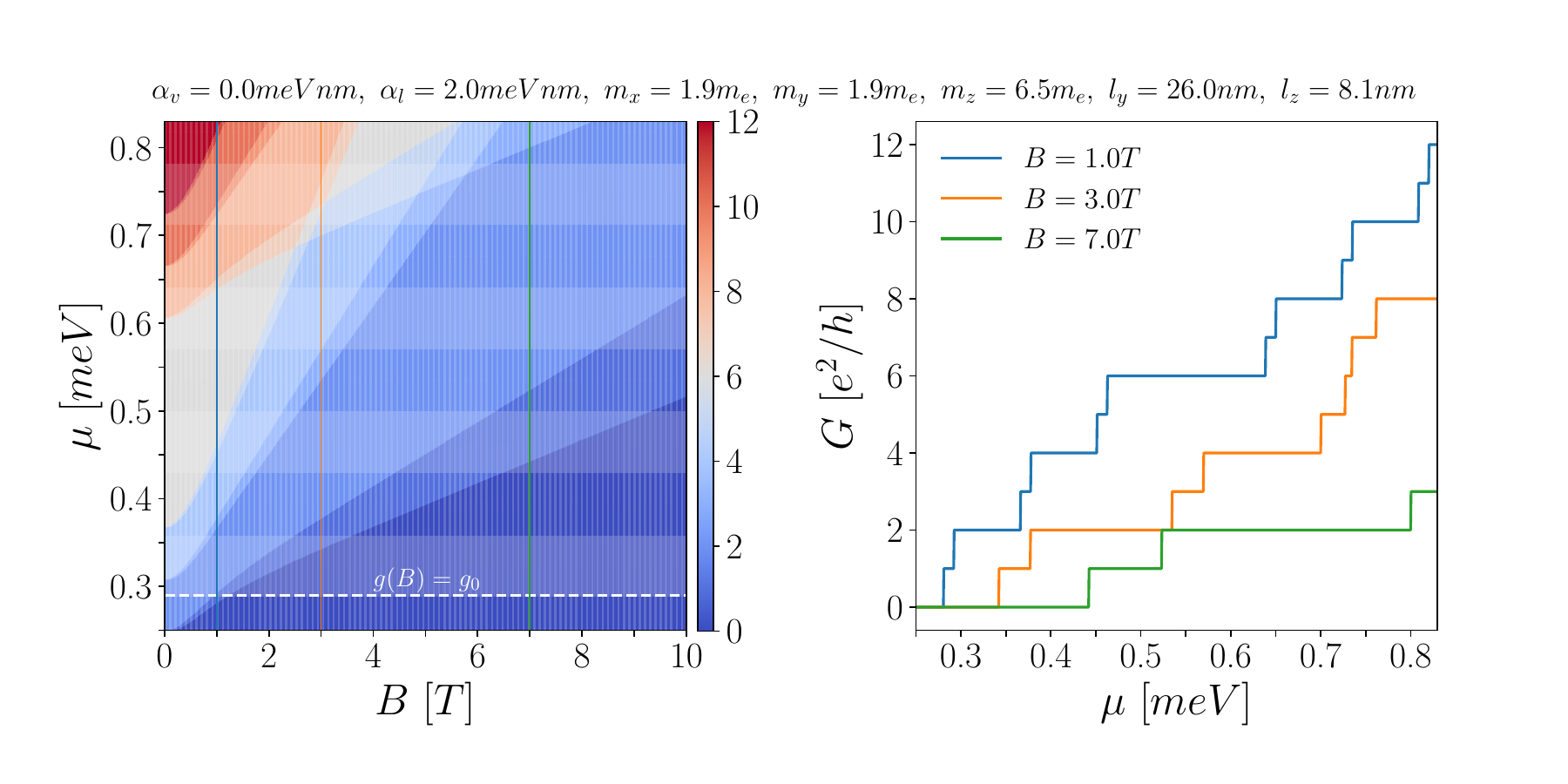}
    
    \caption{(left) Zero-bias conductance $G$ (in units of $e^2/h$) as a function of magnetic field $B$ and chemical potential $\mu$ for $\alpha_v = 0.0$ meV$\,$nm, and $\alpha_l = 2.0$ meV$\,$nm. The dashed white line indicates the renormalized $g$-factor as a guide to the eye, while the three solid vertical lines show the values of $B$ chosen to plot the conductance as a function of $\mu$ in three distinct regimes shown in the right panel. (right) Conductance as a function of $\mu$ for $B = 0.7$\,T (blue), $B = 5$\,T (orange) and $B = 8$\,T. Other parameters are $m_{x} = m_{y} = 1.9m_{e}$ and $m_{z} = 6.5 m_{e}$ where $m_e$ is the electron mass, $l_y = \sqrt{\hslash/(m_y \omega_y)} = 26$ nm, and $l_z = \sqrt{\hslash/(m_z \omega_z)} = 8.1$ nm.}
    \label{fig:lateralSOC}
\end{figure}


\clearpage 

\paragraph{Caption for Movie S1.} 
The supplemental video is an animated version of panels \textbf{A} and \textbf{B} of Figure~\ref{fig:theory_HelicalHarmonicOscillator}, showing the electronic density and the trajectory of the center of mass of the electron along the waveguide in the first and second eigenstate of the chiral potential.


\begin{thebibliography}{10}
\providecommand{\url}[1]{\texttt{#1}}
\expandafter\ifx\csname urlstyle\endcsname\relax
  \providecommand{\doi}[1]{doi:\discretionary{}{}{}#1}\else
  \providecommand{\doi}{doi:\discretionary{}{}{}\begingroup \urlstyle{rm}\Url}\fi

\bibitem{Ray1999-eo}
K.~Ray, Asymmetric Scattering of Polarized Electrons by Organized Organic Films of Chiral Molecules. \emph{Science} \textbf{283}~(5403), 814--816 (1999), \doi{10.1126/science.283.5403.814}, \url{https://dx.doi.org/10.1126/science.283.5403.814}.

\bibitem{Michaeli2016-ei}
K.~Michaeli, N.~Kantor-Uriel, R.~Naaman, D.~H. Waldeck, The electron's spin and molecular chirality - how are they related and how do they affect life processes? \emph{Chemical Society reviews} \textbf{45}~(23), 6478--6487 (2016), \doi{10.1039/c6cs00369a}, \url{https://pubs.rsc.org/en/content/articlelanding/2016/cs/c6cs00369a}.

\bibitem{Naaman2019-wn}
R.~Naaman, Y.~Paltiel, D.~H. Waldeck, Chiral molecules and the electron spin. \emph{Nature Reviews Chemistry} \textbf{3}~(4), 250--260 (2019), \doi{10.1038/s41570-019-0087-1}, \url{https://www.nature.com/articles/s41570-019-0087-1}.

\bibitem{Evers2022-wi}
F.~Evers, \emph{et~al.}, Theory of Chirality Induced Spin Selectivity: Progress and Challenges. \emph{Advanced materials} \textbf{34}~(13), e2106629 (2022), \doi{10.1002/adma.202106629}, \url{http://dx.doi.org/10.1002/adma.202106629}.

\bibitem{Feynman1982-gw}
R.~P. Feynman, Simulating physics with computers. \emph{Intl. J. of Theoretical Physics} \textbf{21}, 467 (1982).

\bibitem{Altman2021-qz}
E.~Altman, \emph{et~al.}, Quantum Simulators: Architectures and Opportunities. \emph{PRX Quantum} \textbf{2}~(1), 017003 (2021), \doi{10.1103/PRXQuantum.2.017003}, \url{https://link.aps.org/doi/10.1103/PRXQuantum.2.017003}.

\bibitem{Greiner2002-ox}
M.~Greiner, O.~Mandel, T.~Esslinger, T.~W. Hansch, I.~Bloch, Quantum phase transition from a superfluid to a Mott insulator in a gas of ultracold atoms. \emph{Nature} \textbf{415}~(6867), 39--44 (2002), \doi{10.1038/415039a}, \url{http://dx.doi.org/10.1038/415039a}.

\bibitem{O-Hara2002-by}
K.~M. O'Hara, S.~L. Hemmer, M.~E. Gehm, S.~R. Granade, J.~E. Thomas, Observation of a strongly interacting degenerate Fermi gas of atoms. \emph{Science (New York, N.Y.)} \textbf{298}~(5601), 2179--2182 (2002), \doi{10.1126/science.1079107}, \url{https://www.science.org/doi/10.1126/science.1079107}.

\bibitem{Bourdel2004-lb}
T.~Bourdel, \emph{et~al.}, Experimental study of the {BEC}-{BCS} crossover region in lithium 6. \emph{Physical review letters} \textbf{93}~(5), 050401 (2004), \doi{10.1103/PhysRevLett.93.050401}, \url{http://link.aps.org/pdf/10.1103/PhysRevLett.93.050401}.

\bibitem{Singha2011-rl}
A.~Singha, \emph{et~al.}, Two-Dimensional Mott-Hubbard Electrons in an Artificial Honeycomb Lattice. \emph{Science} \textbf{332}~(6034), 1176--1179 (2011), \doi{10.1126/science.1204333}, \url{http://dx.doi.org/10.1126/science.1204333}.

\bibitem{Abo-Shaeer2001-vd}
J.~R. Abo-Shaeer, C.~Raman, J.~M. Vogels, W.~Ketterle, Observation of vortex lattices in Bose-Einstein condensates. \emph{Science (New York, N.Y.)} \textbf{292}~(5516), 476--479 (2001), \doi{10.1126/science.1060182}, \url{https://www.science.org/doi/10.1126/science.1060182}.

\bibitem{Lin2009-iu}
Y.~J. Lin, R.~L. Compton, K.~Jimenez-Garcia, J.~V. Porto, I.~B. Spielman, Synthetic magnetic fields for ultracold neutral atoms. \emph{Nature} \textbf{462}~(7273), 628--632 (2009), \doi{10.1038/nature08609}, \url{http://dx.doi.org/10.1038/nature08609}.

\bibitem{Krinner2015-dv}
S.~Krinner, D.~Stadler, D.~Husmann, J.-P. Brantut, T.~Esslinger, Observation of quantized conductance in neutral matter. \emph{Nature} \textbf{517}~(7532), 64--67 (2015), \doi{10.1038/nature14049}, \url{http://dx.doi.org/10.1038/nature14049}.

\bibitem{Briggeman2021-ot}
M.~Briggeman, \emph{et~al.}, One-dimensional {Kronig–Penney} superlattices at the {LaAlO} 3 /{SrTiO} 3 interface. \emph{Nature physics} pp. 1--6 (2021), \doi{10.1038/s41567-021-01217-z}, \url{https://www.nature.com/articles/s41567-021-01217-z}.

\bibitem{Briggeman2020-wp}
M.~Briggeman, \emph{et~al.}, Engineered spin-orbit interactions in {LaAlO3}/{SrTiO3}-based {1D} serpentine electron waveguides. \emph{Science advances} \textbf{6}~(48) (2020), \doi{10.1126/sciadv.aba6337}, \url{http://dx.doi.org/10.1126/sciadv.aba6337}.

\bibitem{Cen2008-ql}
C.~Cen, \emph{et~al.}, Nanoscale control of an interfacial metal–insulator transition at room temperature. \emph{Nature materials} \textbf{7}~(4), 298--302 (2008), \url{https://scholar.google.com/citations?view_op=view_citation&hl=en&user=SOYo5jgAAAAJ&citation_for_view=SOYo5jgAAAAJ:d1gkVwhDpl0C}.

\bibitem{Cen2009-of}
C.~Cen, S.~Thiel, J.~Mannhart, J.~Levy, Oxide nanoelectronics on demand. \emph{Science (New York, N.Y.)} \textbf{323}~(5917), 1026--1030 (2009), \doi{10.1126/science.1168294}, \url{https://www.science.org/doi/10.1126/science.1168294}.

\bibitem{Annadi2018-az}
A.~Annadi, \emph{et~al.}, Quantized Ballistic Transport of Electrons and Electron Pairs in {LaAlO3}/{SrTiO3} Nanowires. \emph{Nano letters} \textbf{18}~(7), 4473--4481 (2018), \doi{10.1021/acs.nanolett.8b01614}, \url{https://doi.org/10.1021/acs.nanolett.8b01614}.

\bibitem{Briggeman2020-cl}
M.~Briggeman, \emph{et~al.}, Pascal conductance series in ballistic one-dimensional {LaAlO3}/{SrTiO3} channels. \emph{Science} \textbf{367}~(6479), 769--772 (2020), \doi{10.1126/science.aat6467}, \url{https://science.sciencemag.org/content/sci/367/6479/769.full.pdf}.

\bibitem{Glazman1989-yr}
L.~I. Glazman, A.~V. Khaetskii, Nonlinear Quantum Conductance of a Lateral Microconstraint in a Heterostructure. \emph{Europhysics letters} \textbf{9}~(3), 263--267 (1989), \doi{10.1209/0295-5075/9/3/013}, \url{http://iopscience.iop.org/article/10.1209/0295-5075/9/3/013/pdf}.

\bibitem{Patel1990-qo}
N.~K. Patel, \emph{et~al.}, Ballistic transport in one dimension: additional quantisation produced by an electric field. \emph{Journal of physics. Condensed matter: an Institute of Physics journal} \textbf{2}~(34), 7247 (1990), \url{http://stacks.iop.org/0953-8984/2/i=34/a=018}.

\bibitem{Damanet2021-lt}
F.~Damanet, \emph{et~al.}, Spin-orbit-assisted electron pairing in one-dimensional waveguides. \emph{Physical review. B, Condensed matter} \textbf{104}~(12), 125103 (2021), \doi{10.1103/PhysRevB.104.125103}, \url{https://link.aps.org/doi/10.1103/PhysRevB.104.125103}.

\bibitem{Mikheev2023-iy}
E.~Mikheev, \emph{et~al.}, A clean ballistic quantum point contact in strontium titanate. \emph{Nature electronics} pp. 1--8 (2023), \doi{10.1038/s41928-023-00981-5}, \url{https://www.nature.com/articles/s41928-023-00981-5}.

\bibitem{Gutierrez2012-zl}
R.~Gutierrez, E.~Díaz, R.~Naaman, G.~Cuniberti, Spin-selective transport through helical molecular systems. \emph{Physical Review B: Condensed Matter and Materials Physics} \textbf{85}~(8), 081404 (2012), \doi{10.1103/PhysRevB.85.081404}, \url{https://link.aps.org/doi/10.1103/PhysRevB.85.081404}.

\bibitem{Eckvahl2023-xj}
H.~J. Eckvahl, \emph{et~al.}, Direct observation of chirality-induced spin selectivity in electron donor-acceptor molecules. \emph{Science (New York, N.Y.)} \textbf{382}~(6667), 197--201 (2023), \doi{10.1126/science.adj5328}, \url{https://www.science.org/doi/10.1126/science.adj5328}.

\bibitem{Cheng2011-ze}
G.~Cheng, \emph{et~al.}, Sketched oxide single-electron transistor. \emph{Nature nanotechnology} \textbf{6}~(6), 343--347 (2011), \doi{10.1038/nnano.2011.56}, \url{https://scholar.google.com/citations?view_op=view_citation&hl=en&user=SOYo5jgAAAAJ&citation_for_view=SOYo5jgAAAAJ:aqlVkmm33-oC}.

\bibitem{Caviglia2010-bp}
A.~D. Caviglia, \emph{et~al.}, Tunable Rashba Spin-Orbit Interaction at Oxide Interfaces. \emph{Physical review letters} \textbf{104}~(12), 126803 (2010), \doi{10.1103/PhysRevLett.104.126803}, \url{http://dx.doi.org/10.1103/PhysRevLett.104.126803}.

\end{thebibliography}
\end{document}